\documentclass[superscriptaddress,twocolumn]{revtex4-2}

\usepackage[dvipsnames]{xcolor}
\usepackage{hyperref}
\hypersetup{
  breaklinks=true,
  colorlinks=true,
  allcolors=BlueViolet,
}

\usepackage{graphicx}
\graphicspath{{./figures/}} 

\usepackage[inline]{enumitem} 
\setlist[enumerate]{label=(\arabic*)} 

\usepackage{amssymb,physics,braket,bm} 
\newcommand{\p}[1]{\left(#1\right)} 
\renewcommand{\sp}[1]{\left[#1\right]} 
\renewcommand{\set}[1]{\left\{#1\right\}} 
\newcommand{\bk}{\braket} 

\newcommand{\ii}{\mathrm{i}\mkern1mu} 
\renewcommand{\d}{\partial} 
\renewcommand{\dd}{\text{d}} 

\newcommand{\x}{\text{x}}
\newcommand{\y}{\text{y}}
\newcommand{\z}{\text{z}}

\newcommand{\C}{\mathcal{C}}
\renewcommand{\H}{\mathcal{H}}
\newcommand{\N}{\mathcal{N}}
\renewcommand{\O}{\mathcal{O}}

\renewcommand{\S}{\mathcal{S}}
\newcommand{\T}{\mathcal{T}}

\newcommand{\EE}{\mathbb{E}}
\renewcommand{\SS}{\mathbb{S}}
\newcommand{\ZZ}{\mathbb{Z}}

\newcommand{\best}{\text{best}}
\newcommand{\worst}{\text{worst}}

\newcommand{\con}{\text{con}}
\newcommand{\init}{\text{init}}
\newcommand{\ramp}{\text{ramp}}
\newcommand{\obj}{\text{obj}}
\newcommand{\tot}{\text{tot}}
\newcommand{\feas}{\text{feas}}
\newcommand{\driver}{\text{driver}}
\newcommand{\eff}{\text{eff}}
\newcommand{\interp}{\text{interp}}
\newcommand{\opt}{\text{opt}}
\newcommand{\prob}{\text{prob}}
\newcommand{\rot}{\text{rot}}
\newcommand{\CD}{\text{CD}}
\newcommand{\tmax}{\text{max}}

\newcommand{\MIS}{\text{MIS}}
\newcommand{\ETF}{\text{ETF}}
\DeclareMathOperator{\NAV}{\text{NAV}}
\DeclareMathOperator{\Price}{\text{Price}}

\usepackage{pifont} 
\newcommand{\yes}{\textcolor{ForestGreen}{\ding{51}}}
\newcommand{\no}{\textcolor{RedOrange}{\ding{55}}}

\newcommand{\JPMC}{
  \affiliation{JPMorganChase, New York City, New York, USA}
}
\newcommand{\Infleqtion}{
  \affiliation{Infleqtion, Chicago, Illinois, USA}
}
\newcommand{\Caltech}{
  \affiliation{Caltech, Pasadena, lifornia, USA}
}
\newcommand{\UChicago}{
  \affiliation{University of Chicago, Chicago, Illinois, USA}
}

\usepackage{orcidlink}

\begin{document}

\title{Q-CHOP: Quantum constrained Hamiltonian optimization}
\author{Michael A.~Perlin\,\orcidlink{0000-0002-9316-1596}}
\email{michael.perlin@jpmchase.com}
\JPMC
\Infleqtion
\author{Ruslan Shaydulin\,\orcidlink{0000-0002-8657-2848}}
\JPMC
\author{Benjamin P.~Hall\,\orcidlink{0000-0002-6238-8675}}
\Infleqtion
\author{Pierre Minssen\,\orcidlink{0000-0003-0361-6962}}
\JPMC
\author{Changhao Li\,\orcidlink{0000-0002-3019-5887}}
\JPMC
\author{Kabir Dubey\,\orcidlink{0000-0002-7621-9004}}
\Infleqtion
\UChicago
\author{Rich Rines\,\orcidlink{0009-0005-4073-1291}}
\Infleqtion
\author{Eric R.~Anschuetz\,\orcidlink{0000-0002-9825-3692}}
\Infleqtion
\Caltech
\author{Marco Pistoia\,\orcidlink{0000-0001-9002-1128}}
\JPMC
\author{Pranav Gokhale\,\orcidlink{0000-0003-1946-4537}}
\Infleqtion


\begin{abstract}
  Combinatorial optimization problems that arise in science and industry typically have constraints.
  Yet the presence of constraints makes them challenging to tackle using both classical and quantum optimization algorithms.
  We propose a new quantum algorithm for constrained optimization, which we call quantum constrained Hamiltonian optimization (Q-CHOP).
  Our algorithm leverages the observation that for many problems, while the best solution is difficult to find, the worst feasible (constraint-satisfying) solution is known.
  The basic idea of Q-CHOP is to enforce a Hamiltonian constraint at all times, thereby restricting evolution to the subspace of feasible states, and slowly ``rotate'' an objective Hamiltonian to trace an adiabatic path from the worst feasible state to the best feasible state.
  Q-CHOP thereby assigns qualitatively distinct roles to the constraint and objective functions of a constrained optimization problem.
  We additionally propose a version of Q-CHOP that can start in \emph{any} feasible state.
  Finally, we benchmark Q-CHOP against the commonly-used adiabatic algorithm of quantum annealing with an objective function that penalizes constraint violation, and find that Q-CHOP consistently performs significantly better on a wide range of problems, including textbook graph problems, knapsack problems, combinatorial auctions, and a real-world financial use case of bond exchange-traded fund basket optimization.
\end{abstract}

\maketitle

\section{Introduction}

Constrained combinatorial optimization problems are ubiquitous in science and industry, with applications including molecular design, resource allocation, and quantitative finance \cite{herman2023quantum, shaydulin2019hybrid, abbas2024challenges}.
The goal of such problems is to find the minimum of an objective function under a specified set of constraints.
The discrete nature of the search space and the presence of constraints pose significant computational challenges.
Quantum computers offer a new paradigm for tackling these challenges, with the promise of breaking both fundamental and practical limits of classical computation.
The ubiquity of constrained optimization problems and the rapid development of quantum computers motivates the study of quantum algorithms as a new approach to constrained optimization.

A number of quantum algorithms have been proposed for constrained optimization.
These include the quantum minimum-finding algorithm of D\"{u}rr and H\o{}yer \cite{durr1999quantum}, the short-path algorithm \cite{hastings2018short, dalzell2023mind}, quantum algorithms for backtracking \cite{montanaro2016quantum, campbell2019applying}, branch-and-bound \cite{montanaro2020quantum}, branch-and-cut \cite{chakrabarti2022universal}, quantum annealing \cite{morita2008mathematical}, quantum-accelerated simulated annealing \cite{somma2008quantum, boixo2015fast}, and the quantum approximate optimization algorithm \cite{hogg2000quantum, farhi2014quantum, hadfield2019quantum}, among others.
A particularly promising class of algorithms is those based on adiabatic quantum computation \cite{farhi2000quantum, albash2018adiabatic}.
Adiabatic quantum computing is an alternative model of quantum computation that is polynomially equivalent to---and can be efficiently simulated by---standard gate-based quantum compuation \cite{aharonov2008adiabatic, yu2018exact}.
In recent years, adiabatic quantum algorithms have been shown to lead to linear systems solvers with optimal scaling \cite{costa2022optimal}, faster algorithms for linear programming \cite{augustino2023quantum}, and continuous optimization algorithms that are competitive with state-of-the-art solvers \cite{leng2023quantum}.
The study of adiabatic quantum algorithms has been further accelerated by the availability of prototype implementations in hardware modalities with hundreds to thousands of qubits \cite{ebadi2022quantum, king2023quantum}.

Adiabatic quantum algorithms solve an optimization problem by following an adiabatic path from an easy-to-prepare ground state of an initial Hamiltonian to the ground state of a final Hamiltonian that encodes the problem.
A common approach to account for constraints when solving an optimization problem with an adiabatic algorithm is to add a penalty for constraint violation to the objective function \cite{zaman2022pyqubo, qiskit2024converters, qubovert2024welcome, alessandroni2025alleviating}.
This penalty drives evolution towards the constraint-satisfying subspace of feasible states.
An alternative approach is to restrict the entire evolution to the in-constraint subspace.
This approach been shown to lead to improved performance on some problems \cite{hen2016quantum}, motivating the development of a general class of techniques that enforce constraints throughout adiabatic evolution.

Though some specialized techniques have been proposed for restricting adiabatic evolution to the in-constraint subspace, general-purpose methods are lacking.
The specialized adiabatic algorithms for finding large independent sets in Refs.~\cite[Section VII]{farhi2014quantum} and \cite{wu2020quantum, yu2021quantum} start the evolution in the computational basis state encoding the smallest possible independent set (an empty set), and use different adiabatic paths to prepare a large independent set.
The quantum state remains in the in-constraint subspace throughout time evolution.
The algorithm in Refs.~\cite{wu2020quantum, yu2021quantum}, in particular, has been shown to be mathematically equivalent to finding a maximum independent set via the Rydberg blockade  \cite{ebadi2022quantum, zhao2025quantum}.
However, the generalization of these techniques to generic constrained optimization problems (and in some cases the efficiency of their implementation \cite{hadfield2019quantum}) remains unclear.
In-constraint subspace mixing has been extended to a general class of constraints in discrete, variational relaxations of continuous-time algorithms \cite{marsh2019quantum, marsh2020combinatorial, tomesh2023divide, saleem2023approaches}, at the cost of losing the connection to the adiabatic theorem due to the difficulty of preparing an initial ground state of the mixing operator.
In addition to being necessary for continuous-time adiabatic algorithms, however, an alignment between initial state and mixing operator has been shown to be important for the performance of variational algorithms \cite{he2023alignment}.
Finally, quantum Zeno dynamics \cite{herman2023constrained} and subspace correction \cite{pawlak2023subspace} were recently proposed as general ways to enforce constraints.
However, these methods require repeated measurements of carefully chosen operators throughout time evolution, which may add considerable overhead.
It is in any case desirable to design unitary algorithms that may be complimentary with measurement-based techniques.

In this work, we extend the idea of adiabatic mixing within the feasible subspace \cite{farhi2014quantum, wu2020quantum, yu2021quantum} to a general framework that is applicable to a broad class of constrained optimization problems.
We refer to our algorithm as quantum constrained Hamiltonian optimization (Q-CHOP).
A key insight for Q-CHOP is that whereas finding the best feasible state of a constrained optimization problem is difficult, finding the worst feasible state is often easy.
The basic idea of Q-CHOP is then to use the objective function of the optimization problem to construct an adiabatic path from the worst feasible state (an easy-to-prepare classical bitstring) to the best feasible state, while enforcing a Hamiltonian constraint at all times to restrict dynamics to the feasible subspace.
We also discuss a relaxation of the requirement to identify and prepare a worst feasible state.
In total, Q-CHOP has two free parameters: (a) a total runtime, which is common to all adiabatic algorithms, and (b) a scalar that is analogous to the penalty factor used in standard approaches to constrained optimization.
Unlike the standard approaches, however, Q-CHOP provides clear guidelines for selecting this penalty factor.
While these free parameters do not allow Q-CHOP to solve \emph{arbitrary} constrained optimization problems, most notably those whose feasible subspace is loosely connected---or disconnected entirely---by the mixer that Q-CHOP constructs out of the objective Hamiltonian, we provide strong numerical evidence that Q-CHOP outperforms the standard adiabatic algorithm for a variety of textbook and industrial constrained optimization problems.

The remainder of this paper is structured as follows.
We briefly review constrained optimization and the standard adiabatic algorithm in Section \ref{sec:background}.
We present an overview of the general strategy for Q-CHOP in Section \ref{sec:general_strategy}, and discuss the simple case of an \textit{odd} objective function and initialization to the worst feasible state in Section \ref{sec:odd}.
We then generalize the strategy in Section \ref{sec:odd} to arbitary objective functions and initialization to arbitrary feasible states in Section \ref{sec:relaxation}, and discuss the treatment of inequality constraints in Section \ref{sec:inequality}.
We provide numerical benchmarking of Q-CHOP on a variety of optimization problems in Section \ref{sec:numerics}, notably including both odd and non-odd objective functions, as well as both equality and inequality constraints.
We leave the investigation of Q-CHOP with arbitrary-feasible-state initialization to future work.
Finally, we discuss conclusions and future directions in Section \ref{sec:discussion}.

\section{Background}
\label{sec:background}

Consider a constrained combinatorial optimization problem of the form
\begin{align}
  \begin{aligned}
    \text{minimize}: ~ & f(\bm x) \\
    \text{subject to}: ~ & C_i(\bm x) = 0 ~\text{for all}~i, \\
    & \bm x \in \set{0,1}^N.
  \end{aligned}
  \label{eq:classical_problem}
\end{align}
Here $f$ is the objective function and $C_i$ are constraints for the decision variables $\bm x$.
We assume without loss of generality that the all constraint functions are nonnegative, which implies that they are minimized by constraint-satisfying (feasible) bitstrings; otherwise, we can replace $C_i(\bm x)$ in Eq.~\eqref{eq:classical_problem} by $\widetilde C_i(\bm x) = C_i(\bm x)^2$.
We discuss the treatment of inequality constraints in Section \ref{sec:inequality}.

Under the assumption of nonnegative constraints, the optimization problem in Eq.~\eqref{eq:classical_problem} can be encoded into a system of qubits by the objective and constraint Hamiltonians
\begin{align}
  H_\obj = \sum_{\bm x} f(\bm x) \op{\bm x},
  &&
  H_\con = \sum_{i,\bm x} C_i(\bm x) \op{\bm x},
  \label{eq:encoding}
\end{align}
thereby associating ``good'' bitstrings of low objective-function value with low-energy states of $H_\obj$, and identifying the set of feasible bitstrings with the ground-state subspace of $H_\con$.
We note that Eq.~\eqref{eq:encoding} is only a formal expansion of $H_\obj$ and $H_\con$, which are in practice typically expressed as sums of (polynomially many) Pauli strings.

Thus encoded, the optimization problem in Eq.~\eqref{eq:classical_problem} can be solved via the adiabatic algorithm \cite{albash2018adiabatic}.
To this end, the standard procedure is to first construct a modified objective Hamiltonian $H_\obj + \lambda H_\con$ with a penalty factor for constraint violation, $\lambda$, that is sufficiently large to ensure that the ground state of $H_\obj + \lambda H_\con$ is feasible.
One then chooses a \emph{driver} Hamiltonian $H_\driver$ that satisfies two properties:
\begin{enumerate*}[label=(\roman*)]
\item\label{cond:easy} it has an easy-to-prepare ground state, and
\item\label{cond:comm} it does not commute with $H_\obj + \lambda H_\con$.
\end{enumerate*}
Finally, the quantum algorithm prepares the ground state of $H_\driver$, and slowly changes the Hamiltonian of the system from $H_\driver$ to $H_\obj + \lambda H_\con$ over the course of a total time $T$, for example via the linearly interpolating Hamiltonian \cite{albash2018adiabatic}
\begin{align}
  H_\interp(s) = (1-s) H_\driver + s (H_\obj + \lambda H_\con),
  \label{eq:H_interp}
\end{align}
sweeping $s:0\to 1$ as the time $t:0\to T$.
We refer to $T$ as the ``quantum runtime'' of the algorithm.
Condition \ref{cond:easy} is necessary to ensure that the algorithm solves the optimization problem by adiabatically tracking the ground state of $H_\interp$, while condition \ref{cond:comm} ensures that $H_\interp$ has off-diagonal terms that induce nontrivial dynamics.
Choosing an interpolating Hamiltonian of the form in Eq.~\eqref{eq:H_interp} is not required, and indeed sometimes different paths from $H_\driver$ to $H_\prob$ can improve performance, in the extreme case turning algorithmic failure into success \cite{farhi2002quantum}.
For simplicity, however, we will consider the linearly interpolating Hamiltonian in this work, and refer to this approach as the standard quantum adiabatic algorithm (SQAA).

The adiabatic theorem \cite{albash2018adiabatic} ensures that the SQAA prepares an optimal solution in a time $T$ that scales with an inverse power of the smallest spectral gap $\Delta_\interp^{\text{min}}$ of $H_\interp$ (minimized over $s\in[0,1]$), provided that there are no symmetries or selection rules that fragment Hilbert space into uncoupled sectors.
However, the gap $\Delta_\interp^{\text{min}}$ is usually not known in advance, so in practice the SQAA requires making an ad hoc choice of some ``long'' time $T$, and hoping that $T$ is sufficiently large.
If the guessed time is too small, the SQAA will not only fail to provide an optimal solution, it may even fail to provide a \emph{feasible} solution.

There are a few additional shortcomings of the SQAA that are worth noting.
First, the SQAA requires making a careful choice of the penalty factor $\lambda$.
Increasing $\lambda$ makes it more likely that the SQAA finds feasible solutions to the optimization problem at hand.
However, doing so also increases the runtime of the algorithm, or equivalently the depth of a quantum circuit that implements this algorithm with fixed error.
Making an optimal choice of $\lambda$ has been shown to be a difficult task in practice \cite{herman2023constrained, niroula2022constrained, saleem2023approaches} and NP-hard in general \cite{alessandroni2025alleviating}, although heuristic strategies may work well in certain cases \cite{hao2022exploiting, alessandroni2025alleviating}.
Second, the SQAA relies on an ad hoc initial state and driver Hamiltonian $H_\driver$ that has no connection to the problem of interest.
A common choice is to implement quantum annealing \cite{morita2008mathematical} by initializing a uniform superposition over all bitstrings (representing the uniform prior), and using a transverse magnetic field (for which the uniform superposition is a unique ground state) as the driver Hamiltonian.
However, even here one must choose an overall normalization (prefactor) for $H_\driver$, analogous to the choice of penalty factor $\lambda$.
Finally, there is an intuitive sense in which the SQAA does ``too much work'': it searches the configuration space of \emph{all} bitstrings, feasible or otherwise.

\section{Quantum Constrained Hamiltonian Optimization}
\label{sec:QCHOP}

We now introduce our algorithm and describe its application to problems with various objectives and constraints.
Throughout this paper, we will denote the Pauli-$Z$ operator for qubit $j \in \ZZ_N = \set{1,2,\cdots,N}$ by $Z_j = \op{0}_j - \op{1}_j$, the Pauli-$X$ operator by $X_j = \op{0}{1}_j + \op{1}{0}_j$, and the Pauli-$Y$ operator by $Y_j = -\ii Z_j X_j$.
We will also denote the generator of a global rotation about axis $\alpha\in\set{\x,\y,\z}$ on the Bloch sphere by $S_\alpha$, for example $S_\z = \frac12 \sum_j Z_j$, such that $R_\alpha(\theta) = e^{-\ii\theta S_\alpha}$ rotates all qubits about the axis $\alpha$ by the angle $\theta$.

\subsection{General strategy}
\label{sec:general_strategy}

The general strategy for Q-CHOP is as follows:
\begin{enumerate}
  \item\label{step:init} Identify an initial Hamiltonian $H_\init$ whose energy within the space $\H_\feas$ of feasible states is minimized by an easy-to-prepare state $\ket{\bm x_\init}\in\H_\feas$.
    Prepare $\ket{\bm x_\init}$.
  \item Construct a ramp $H_\ramp(s)$ that traces a continuous path from $H_\init$ to the objective Hamiltonian $H_\obj$, with $H_\ramp(0)=H_\init$ and $H_\ramp(1)=H_\obj$.
  \item\label{step:evolve} Pick a time $T$ and a positive real number $\lambda$, and evolve from time $t=0$ to $t=T$ under the total Hamiltonian
    \begin{align}
      H_\tot(s) = H_\con + \lambda^{-1} H_\ramp(s),
    \end{align}
    with $s:0\to 1$ as $t:0\to T$.
\end{enumerate}
We construct an explicit ramp $H_\ramp(s)$ in Sections \ref{sec:odd} and \ref{sec:relaxation} for optimization objectives of increasing generality, but first take some time to discuss general features of the strategy in steps \ref{step:init}--\ref{step:evolve} and related considerations.

In words, as with any adiabatic algorithm Q-CHOP initializes to the ground state of some initial Hamiltonian.
Q-CHOP then restricts dynamics to the feasible subspace $\H_\feas$ with the constraint Hamiltonian $H_\con$, and adiabatically transvers the initial state to an optimal (or approximately optimal) solution to the optimization problem by tracking the in-constraint ground state of a suitably-constructed adiabatic ramp $H_\ramp(s)$.
The restriction to the feasible subspace effectively reduces the size of the classical search space of the quantum algorithm, which can reduce requisite runtimes \cite{hen2016quantum}.

The scalar $\lambda$ is analogous to the penalty factor in the SQAA, and indeed the Hamiltonians of Q-CHOP and the SQAA are equal at $s=1$ (up to an arbitrary overall normalization factor).
However, the total Hamiltonian of Q-CHOP is built entirely from problem data, without reference to an auxiliary driver Hamiltonian.
Whereas in the SQAA, the penalty factor $\lambda$ is used to modify the objective function (that is, only the \emph{final} Hamiltonian in an adiabatic ramp), in Q-CHOP the penalty factor $\lambda$ is used to enforce constraints \emph{at all times}.
Q-CHOP thereby assigns qualitatively distinct roles to the constraint and objective functions of a constrained optimization problem, and moreover provides a theoretical framework for making a suitable choice of $\lambda$.

Denote the operator norm of $M$ by $\norm{M}$, let $\norm{H_\ramp} = \max_{s\in[0,1]} \norm{H_\ramp(s)}$, and denote the spectral gap of $H_\con$ by $\Delta_\con$.
If
\begin{enumerate}[label=(\alph*)]
  \item\label{cond:gap} $\norm{H_\ramp}/\lambda < \Delta_\con/2$, and
  \item\label{cond:slow} $H_\ramp$ changes slowly with respect to $\Delta_\con^{-1}$ \cite{albash2018adiabatic},
\end{enumerate}
then second-order perturbation theory guarantees that Q-CHOP will yield feasible states with probability
\begin{align}
  p_\feas = 1 - O(\delta^2),
  &&
  \text{where}
  &&
  \delta = \frac{\norm{H_\ramp}}{\lambda\Delta_\con}.
  \label{eq:p_feas}
\end{align}
Intuitively, condition \ref{cond:gap} ensures that the ramp $H_\ramp$ is not ``strong enough'' to violate constraints, while condition \ref{cond:slow} precludes rapid (e.g., Floquet) driving schemes that resonantly pump energy into the system.

Larger values of penalty factor $\lambda$ provide better guarantees of feasible solutions, while smaller values of $\lambda$ are desirable to increase the magnitude of the ramp $H_\ramp$ that drives adiabatic evolution within the feasible subspace.
Condition \ref{cond:gap} and the guarantee in Eq.~\eqref{eq:p_feas} provide an upper bound on ``reasonable'' choices of $\lambda$, namely $\lambda_{\text{max}}\approx 2\norm{H_\ramp}/\Delta_\con$
\footnote{
  In practice, even the upper bound $\lambda\le\lambda_{\text{max}}$ is too stringent, as the above discussion still holds if the norm $\norm{H_\ramp(s)}$ is replaced by $\norm{(1-P_\feas) H_\ramp(s) P_\feas}$, where $P_\feas$ is a projector onto the feasible subspace.
  One should therefore expect smaller values of $\lambda$ to still yield predominantly feasible states, although the limits on $\lambda$ may need to be found empirically because the norm $\norm{(1-P_\feas) H_\ramp(s) P_\feas}$ is generally not known a priori.
}.
The ramps that we construct below (Sections \ref{sec:odd} and \ref{sec:relaxation}) have the feature that $\norm{H_\ramp} \le \norm{H_\obj}$, where the objective norm $\norm{H_\obj}$ is typically straightforward to compute or bound from above
\footnote{
  Specifically, if the objective is expanded into a sum of Pauli strings as $H_\obj = \frac12 \sum_{S\subset\ZZ_N} c_S \prod_{j\in S} Z_j$, then its operator norm is bounded as $\norm{H_\obj}\le \frac12 \sum_{S\subset\ZZ_N} \abs{c_S}$.
}.
Meanwhile, the constraint Hamiltonian $H_\con$ typically has a natural expression as a sum of projectors, in which case $\Delta_\con\ge 1$.
A homogeneous linear objective $H_\obj = S_\z = \frac12 \sum_j Z_j$, for example, has norm $\norm{H_\obj} = N/2$, so if $\Delta_\con = 1$ then $\lambda = N$ is a reasonable choice of penalty factor.
The fact that a suitable choice of $\lambda$ can be determined from efficiently computable problem data is a notable advantage of Q-CHOP over the SQAA.
In addition to providing a guarantee of restriction to the feasible subspace, having a natural choice of $\lambda$ potentially avoids the need for a costly hyperparameter tuning routine.
Of course, additional optimization over choices of $\lambda$ would further improve Q-CHOP performance.
As in the SQAA, however, determining the \emph{optimal} choice of $\lambda$ may be NP-hard in general \cite{alessandroni2025alleviating}.

\subsection{Odd objectives and bad solutions}
\label{sec:odd}

It is often the case that the objective function of an optimization problem is linear in the decision variables of the problem, in which case the corresponding objective Hamiltonian can be written (up to an irrelevant constant shift) as a weighted sum of single-qubit Pauli-$Z$ operators.
This Hamiltonian is then \emph{odd} with respect to qubit inversion, which is to say that $\bk{\bm x|H_\obj|\bm x}=-\bk{\bar{\bm x}|H_\obj|\bar{\bm x}}$, where $\bm x = (x_1,x_2,\cdots)\in\set{0,1}^N$ is a bitstring and $\bar{\bm x}=(1-x_1,1-x_2,\cdots)$ is its bitwise complement.
It is also sometimes the case that finding good solutions to a constrained optimization problem is difficult, but finding the \emph{worst} feasible solution is easy, either by inspection or with an efficient classical algorithm.
This is the case for the problem of finding a maximum independent set (MIS), whose worst feasible state is the all-0 bitstring (see Section \ref{sec:MIS}).
When the objective is odd, the asymmetry between best and worst feasible states motivates the choice of initial Hamiltonian $H_\init = -H_\obj$, and a ramp that continuously flips all qubits: $H_\ramp(\theta) = -H_\obj(\theta)$, with
\begin{align}
  H_\obj(\theta) = R_\y(\theta) H_\obj R_\y(\theta)^\dag
  \label{eq:H_obj_rot_odd}
\end{align}
where we have switched to parameterising the ramp by the angle $\theta=s\pi$ for brevity, and $R_\y(\theta) = e^{-\ii\theta S_\y}$ rotates all qubits about the $y$ axis by $\theta$.

When these two conditions hold, namely that objective Hamiltonian is odd and the worst feasible state $\ket{\bm x_\worst}$ is easy to prepare, the Q-CHOP algorithm is as follows.
Prepare $\ket{\bm x_\worst}$, pick a quantum runtime $T$ and positive number $\lambda\approx 2 \norm{H_\obj} / \Delta_\con$, and evolve from time $t=0$ to $t=T$ under the total Hamiltonian
\begin{align}
  H_\tot(\theta) = H_\con - \lambda^{-1} H_\obj(\theta),
  \label{eq:H_tot}
\end{align}
sweeping $\theta:0\to\pi$ as $t:0\to T$, for example with the linear ramp $\theta = \pi t/T$.
Here $\norm{H_\obj}$ is the operator norm of $H_\obj$, and $\Delta_\con$ can be replaced by an estimate or lower bound on the spectral gap of $H_\con$.
The prescription for choosing $\lambda$ is only a guideline, and smaller values of $\lambda$ may work well in practice.
After appropriately normalizing $H_\obj$ (see Section \ref{sec:setup}), we set $\lambda$ equal to the qubit number $N$ in all examples considered in this work.

In the case of MIS, we show in Appendix \ref{sec:equiv} that Q-CHOP is equivalent to the specialized algorithm based on non-Abelian adiabatic mixing proposed in Ref.~\cite{yu2021quantum}, up to an adiabatically vanishing addition of $\d_t\theta S_\y$ to $H_\tot(\theta)$ in Eq.~\eqref{eq:H_tot}.
However, the general formalism of Q-CHOP extends to more general constrained optimization problems as well.
Intriguingly, the addition of $\d_t\theta S_\y$ can be identified with approximate counterdiabatic driving (see Appendix \ref{sec:counterdiabatic}), and its form depends only on the linearity of the MIS objective function.
An added term of $\d_t\theta S_\y$ is therefore expected to marginally improve the performance of Q-CHOP for any optimization problem with linear objectives, and similar approximate counterdiabatic driving terms can likely be derived for non-linear objectives as well.
For simplicity, we leave the consideration of counterdiabatic driving in Q-CHOP to future work, but remark that there may generally be benefits to Q-CHOP from borrowing related shortcut-to-adiabaticity techniques \cite{guery-odelin2019shortcuts, sels2017minimizing, li2024quantum}.

\subsection{Arbitrary objectives and feasible states}
\label{sec:relaxation}

We now relax the requirement of an odd objective used in Sec.~\ref{sec:odd} and discuss a modification of the algorithm that no longer requires the worst feasible state to be easy to prepare.
In the case of a non-odd objective, the prescription in Eq.~\eqref{eq:H_obj_rot_odd} needs modification to ensure that the rotated objective Hamiltonian changes sign when $\theta:0\to\pi$.
To this end, we decompose the objective into Pauli strings as
\begin{align}
  H_\obj = \frac12 \sum_{S\subset\ZZ_N} c_S \prod_{j\in S} Z_j.
  \label{eq:H_obj_z}
\end{align}
We then rotate one qubit in each term, and average over each choice of qubits, altogether defining
\begin{align}
  H_\obj(\theta) = \frac12 \sum_{S\subset\ZZ_N} \frac{c_S}{\abs{S}}
  \sum_{j\in S} R_\y(\theta) Z_j R_\y(\theta)^\dag \prod_{k\in S\setminus\set{j}} Z_k,
  \label{eq:H_obj_rot_gen}
\end{align}
where
\begin{align}
  R_\y(\theta) Z_j R_\y(\theta)^\dag = \cos(\theta) Z_j + \sin(\theta) X_j.
\end{align}
While time evolution by a Hamiltonian of the form in Eq.~\eqref{eq:H_obj_rot_gen} is likely difficult in the setting of analog quantum computation, in a gate-based setting the only added cost over time-evolving by $H_\obj$ comes from the fact that there are more locally-rotated Pauli strings in $H_\obj(\theta)$ than there are Pauli-$Z$ strings in $H_\obj$.
Specifically, for $k$-body objective Hamiltonians the number of terms increases by a factor of $O(k)$.
An alternative strategy is to decompose the objective into even and odd terms, rotate the odd terms globally as in Eq.~\eqref{eq:H_obj_rot_odd}, and rotate the even terms ``locally'' as in Eq.~\eqref{eq:H_obj_rot_gen}.
This alternative strategy may have the benefit of reduced computational overheads when discretizing Q-CHOP for gate-based quantum computation.
We leave the exploration of different strategies to continuously rotate $H_\obj\to-H_\obj$ to future work.

The second nominal requirement of Q-CHOP is that the worst feasible state of a constrained optimization problem is easy to prepare.
In fact, this requirement is satisfied by all problems that we will consider in Section \ref{sec:numerics}.
Nonetheless, here we sketch out an idea to relax this requirement to that of having \emph{at least one} feasible state that is easy to prepare.
Let $\ket{\bm x_\star}$ be an easy-to-prepare eigenstate of $H_\obj$, and let $E_\star = \bk{\bm x_\star|H_\obj|\bm x_\star}$ be the corresponding objective energy.
Define a subproblem with the same constraint Hamiltonian as the original optimization problem, $H_\con$, but a new objective Hamiltonian
\begin{align}
  H_\obj^\star = -(H_\obj - E_\star)^2,
\end{align}
for which $\ket{\bm x_\star}$ is the worst feasible state.
By construction, $H_\obj^\star$ is minimized within the feasible subspace by either the best feasible state $\ket{\bm x_\best}$ or worst feasible state $\ket{\bm x_\worst}$ of the original optimization problem, depending on which corresponding objective energy (with respect to $H_\obj$) is farther from $E_\star$.
If the optimization subproblem defined by $(H_\obj^\star, H_\con)$ is solved by $\ket{\bm x_\best}$, then after solving the subproblem we are done.
Otherwise, we can use the worst feasible state $\ket{\bm x_\worst}$ identified by the subproblem to solve the original optimization problem $(H_\obj, H_\con)$.
The relaxation of Q-CHOP from needing the worst feasible state to needing any feasible state comes at the cost of having to solve a subproblem with an objective $H_\obj^\star$ that has quadratically larger norm, and can therefore be expected to require a larger penalty factor $\lambda^\star$ and a larger quantum runtime $T^*$.
There may additionally be concerns if the energy $E_\star$ has degeneracy $d_\star\gg1$, in which case a population transfer into the ground state of the final Q-CHOP Hamiltonian can be expected to be suppressed by a factor of $1/d_\star$.
We leave an investigation of these and other considerations that may affect the viability of the worst-to-any feasible state relaxation to future work.

\subsection{Inequality constraints}
\label{sec:inequality}

Here we discuss the treatment of inequality constraints in Q-CHOP and the SQAA.
An inequality constraint of the form $D(\bm x)\ge 0$ can be converted into an equality constraint by the introduction of a \emph{slack variable} $n\in\set{0, 1, 2, \cdots}$ that is constrained to satisfy $n = D(\bm x)$ by the constraint
\begin{align}
  C_D(\bm x,n) = (D(\bm x) - n)^2 = 0.
\end{align}
The Hamiltonian enforcing this constraint takes the form
\begin{align}
  H_\con^D = \p{\sum_{\bm x} D(\bm x) \op{\bm x} - \hat n_D}^2,
  \label{eq:H_con_ineq}
\end{align}
where $\hat n_D = \sum_{n\ge0} n \op{n}_D$ is a number operator that can be identified with an auxiliary bosonic mode indexed by the constraint function $D$.
In this sense, bosonic modes may be resources for enforcing inequality constraints.

In practice, the dimension of a slack variable can be truncated at $D_\tmax = \max_{\bm x} D(\bm x)$ to encode a problem into a finite-dimensional Hilbert space.
A discrete-variable encoding of the slack variable then requires $\lceil\log_2(D_\tmax+1)\rceil$ qubits, though it may be desirable to encode the slack variable directly into a qudit or oscillator \cite{gottesman2001encoding, noh2020encoding}.
If the maximum of $D$ is not efficiently computable, $D_\tmax$ can be replaced by an upper bound on $D(\bm x)$.
Expanding
\begin{align}
  D(\bm x) = \sum_{S\subset \ZZ_N} c_S^D \prod_{j\in S} x_j,
\end{align}
where each $x_j\in\set{0, 1}$, we denote the constant contribution to $D(\bm x)$ by $c_0^D = c_{\set{}}^D$, and the set of remaining coefficients by $\C_D = \set{c_S^D:S\subset\ZZ_N,\abs{S}>0}$.
A suitable upper bound for $D(\bm x)$ is then given by the sum of positive coefficients in $\C_D$,
\begin{align}
  \widetilde D_\tmax = \mathop{\mathrm{sum}}(\C_D\cap\ZZ_+) + c_0^D,
\end{align}
where $\ZZ_+ = \set{0, 1, 2, \cdots}$.
If the coefficients in $\C_D$ have a nontrivial greatest common divisor $\mathop{\mathrm{gcd}}(\C_D)>1$, the dimensionality of the slack variable can be further reduced by pruning its allowed values to the set
\begin{align}
  \SS_D = (\mathop{\mathrm{gcd}}(\C_D) \ZZ + c_0^D)
  \cap \sp{0, \widetilde D_\tmax}.
\end{align}

For the SQAA, the introduction of a slack variable forces a choice of initial state and driver Hamiltonian for the corresponding slack qudit.
A natural generalization of the qubit-only case is to interpret the initial state $\ket{+}^{\otimes N}$ and transverse-field driver Hamiltonian $H_\driver = -\frac12\sum_j X_j \simeq -\sum_j \op{+}_j$ respectively as a uniform prior and a sum of projectors onto the uniform prior for each decision variable.
Here $\simeq$ denotes equality up to an overall constant.
This interpretation motivates initializing the slack qudit into a uniform superposition of its domain, and adding the (negative) projector onto this state to the driver Hamiltonian.
We will adopt this generalization of initial state and driver Hamiltonian in this work, and leave the study of alternative generalizations for qudits (and infinite-dimensional bosonic modes) to future work.

In the case of Q-CHOP, there are two subtleties to address for the treatment of inequality constraints.
First, once an initial worst feasible state $\ket{\bm x_\worst}$ is identified, the slack variable for inequality constraint $D$ should be initialized to $\ket{n_\worst} = \ket{D(\bm x_\worst)}$.
Second, merely adding a slack qudit with the constraint in Eq.~\eqref{eq:H_con_ineq} to the Q-CHOP Hamiltonian fragments Hilbert space into uncoupled sectors indexed by the value of $n=D(\bm x)$.
Initializing the slack variable in $\ket{n_\worst}$ thereby enforces, in effect, the much stronger constraint that $D(\bm x) = n_\worst$.
To remedy this situation, we borrow the strategy of engineering an alternative adiabatic path \cite{farhi2002quantum} that adds thoughtfully chosen mixing terms to the Q-CHOP Hamiltonian.

In order to satisfy the constraint enforced by Eq.~\eqref{eq:H_con_ineq}, our added mixing terms need to couple the decision variables and the slack variables, changing them in a correlated manner.
A Hamiltonian term that flips qubit $j$, for example, should simultaneously change the value of all slack variables associated with inequality constraints that involve variable $x_j$, so as to ensure that inequality constraints remain satisfied.
However, engineering a strictly constraint-satisfying mixing Hamiltonian would involve complex multi-control logic that may generally be inefficient to implement.
Instead, we can allow slack variables to change to \emph{any} value, and rely on the always-on constraint Hamiltonian of Q-CHOP to make constraint-violating terms off-resonant.
For problems with inequality constraints, we therefore modify the objective Hamiltonian of Q-CHOP as $H_\obj(\theta) \to H_\obj(\theta) \times \S(\theta)$, with
\begin{align}
  \S(\theta) = 1 + \sin(\theta) \prod_D \T_D,
  &&
  \T_D = \sum_{m,n\in\SS_D} \op{m}{n}_D.
\end{align}
The factor of $\sin\theta$ ensures that the added mixing terms vanish at $\theta=0$ and $\pi$, which is necessary to preserve the initial and final ground states of the Q-CHOP Hamiltonian.

The mixing operator $\T_D$ can be written as as $\T_D = \abs{\SS_D} P_D^+$, where $\abs{\SS_D}$ is the dimension of the slack qudit for constraint $D$, and $P_D^+ = \op{+}{+}_D$ is the projector onto the uniform superposition $\ket{+}_D \propto \sum_{n\in\SS_D} \ket{n}_D$.
If the polynomial $D(\bm x)$ has degree $d$ and coefficients with magnitude $O(1)$, then $\norm{\T_D} = \abs{\SS_D} = O(N^d)$.
It is therefore worth noting that the norm $\norm{\S(\theta)}$ grows with the product of the slack variable dimensions, $\prod_D \abs{\SS_D}$.
Though it is not surprising for the cost of solving an optimization problem to grow exponentially with the degree of its inequality constraints (e.g., with quadratic constraints inducing a quadratic cost), it is highly desirable to have a cost that does not grow multiplicatively with the \emph{number} of inequality constraints.
In future work, it is therefore imperative to find alternative strategies to enforce inequality constraints and engineer new adiabatic paths through the induced subspace of feasible states.

\section{Numerical experiments}
\label{sec:numerics}

We now study the performance of Q-CHOP numerically using exact classical simulations, and compare it to the SQAA.
A comprehensive performance comparison between Q-CHOP and the SQAA requires a detailed, problem-aware assessment of the numerous choices that are made to instantiate simulation Hamiltonians, including choices of normalization factors, driver Hamiltonians (for the SQAA), penalty factor, adiabatic schedules, and total evolution times.
Since an exhaustive study these hyperparameters is infeasible computationally, in this work we use physically-motivated choices that allow for a high-level assessment of Q-CHOP as a general-purpose strategy for constrained combinatorial optimization.

\subsection{Experimental Setup}
\label{sec:setup}

To simulate Q-CHOP, we evolve an initial state with the Hamiltonian in Eq.~\eqref{eq:H_tot} using a linear ramp, $\theta=\pi t/T$.
In the case of the SQAA, we simulate quantum annealing by evolving the initial state $\ket{+}^{\otimes N}\propto(\ket{0}+\ket{1})^{\otimes N}$ with the Hamiltonian
\begin{align}
  H_{\text{SQAA}}(t) = -\p{1-\frac{t}{T}} S_\x + \frac{t}{T} (H_\con + \lambda^{-1} H_\obj),
\end{align}
where $S_\x = \frac12 \sum_j X_j \simeq \sum_j \op{+}{+}_j$ is a transverse field normalized to have a spectral range equal to the qubit number $N$.
This field was chosen as the simplest driver Hamiltonian that allows for quantum annealing from the uniform prior $\ket{+}^{\otimes N}$.
For optimization problems with slack qudits, we initialize the qudits to a uniform superposition and add projectors onto these initial states to $S_\x$.%
Reflecting the analogous roles played by the penalty factor $\lambda$ in the SQAA and Q-CHOP, we set $\lambda=N$ in all cases, such that the final Hamiltonians of the SQAA and Q-CHOP are equal.
These methods thereby differ only in the adiabatic paths that they take to arrive at the same final Hamiltonian.
We defer a detailed numerical study of Q-CHOP and SQAA performance when optimizing $\lambda$ to future work.

In all experiments, we normalize the objective Hamiltonian to make the optimization problems scale-invariant.
Specifically, we set $H_\obj \to H_\obj/\N(H_\obj)$, where the normalization factor $\N(H_\obj)$ is defined using the nonzero coefficients of $H_\obj$ in Eq.~\eqref{eq:H_obj_z} by
\begin{align}
  \N(H_\obj)^2 = \mathop{\EE}_{\substack{c_S\ne0\\S\ne\set{}\\}} c_S^2.
\end{align}
Here $\EE$ denotes an average, and we exclude the empty set to make the norm $\N(H_\obj)$ independent of constant shifts to $H_\obj$.
We note that this normalization is equivalent to that of Ref.~\cite{sureshbabu2024parameter} up to a constant factor of four.
This factor is chosen so that $\N(S_\z) = 1$ for the homogeneous linear objective $S_\z \simeq -\sum_j\op{1}_j$ with spectral range $N$.

Unlike the objective Hamiltonians, constraint Hamiltonians typically have a natural expression as a sum of projectors, each of which has a spectral range of 1.
Constraint Hamiltonians thereby have a spectral gap of at least 1, and typically equal (or otherwise close) to 1.
We therefore perform no additional preconditioning of the constraint Hamiltonians unless specified otherwise.
For problems with inequality constraints, we divide all constraint coefficients by their greatest common divisor by taking $D(\bm x)\to D(\bm x)/\mathop{\mathrm{gcd}}(\C_D\cup c_0^D)$, prune the set of allowed slack variable values, and modify the SQAA and Q-CHOP Hamiltonians to address slack variables as discussed in Section \ref{sec:inequality}.

We use two metrics to measure the performance of an optimization algorithm.
The first is the in-constraint approximation ratio, which is defined such that the best and worst feasible states have values of 1 and 0, respectively.
The operator whose expectation is equal to the in-constraint approximation ratio is given by
\begin{align}
  r = P_\feas \p{\frac{H_\obj - E_\worst}{E_\best - E_\worst}} P_\feas,
\end{align}
where $P_\feas$ is a projector onto the feasible-state manifold, and $E_\ell = \bk{\bm x_\ell|H_\obj|\bm x_\ell}$ for $\ell\in\set{\best,\worst}$ are the objective energies of the best and worst feasible states.

Our second performance metric is the probability with which measuring a state in the computational basis yields an optimal solution to the optimization problem.
This probability is measured by the projector onto the space of optimal solutions,
\begin{align}
  P_\opt = \sum_{\bm x:\bk{\bm x|r|\bm x} = 1} \op{\bm x}.
\end{align}

For optimization problems whose objective function values may lie anywhere on the real line, finding the exact solution is too strict of a criterion for success.
In this case, we instead report the overlap with $\epsilon$-optimal states, measured by the projector
\begin{align}
  P_\epsilon = \sum_{\bm x:\bk{\bm x|r|\bm x} \ge 1 - \epsilon} \op{\bm x}.
\end{align}

Table \ref{tab:problems} summarizes the problems that we simulate in our numerical experiments, and some features that are relevant for Q-CHOP.
All simulations were performed by numerically integrating equations of motion using the \texttt{DOP853} method of \texttt{scipy.integrate.solve\_ivp}, with both absolute and relative error tolerances of $10^{-8}$ \cite{virtanen2020scipy}.

\begin{table}
  \centering
  \begin{tabular}{cl|c|c}
    \multicolumn{2}{c|}{problem}                      &
    \begin{tabular}{c} odd \\ objective
    \end{tabular} &
    \begin{tabular}{c} no inequality \\ constraints
    \end{tabular}                   \\ \hline\hline
    (\ref{sec:MIS})                                   & MIS           & \yes & \yes \\ \hline
    (\ref{sec:DMDS})                                  & DMDS          & \yes & \yes \\ \hline
    (\ref{sec:KS})                                    & knapsack      & \yes & \no  \\ \hline
    (\ref{sec:CA})                                    & comb.~auction & \yes & \no  \\ \hline
    (\ref{sec:ETF})                                   & ETF           & \no  & \no
  \end{tabular}
  \caption{Problems simulated in Section \ref{sec:numerics}, and features that are relevant for Q-CHOP.}
  \label{tab:problems}
\end{table}

\subsection{Maximum independent set}
\label{sec:MIS}

\begin{figure*}
  \centering
  \includegraphics[scale=0.75]{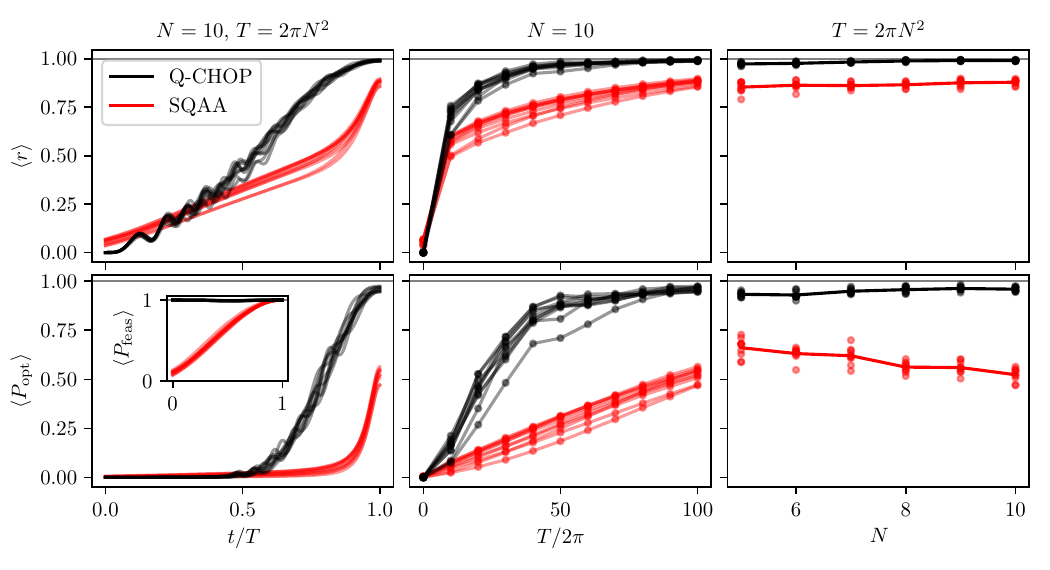}
  \caption{
    Summary of MIS simulation results with Q-CHOP and the SQAA.
    (\textbf{Top row}) In-constraint approximation ratio $\bk{r}$ throughout simulations.
    (\textbf{Top row}) Expectation value of the optimal-state projector $P_\opt$ throughout simulations.
    (\textbf{Left-most panels}) Time-series data in simulations of 10 random graphs with $N=10$ vertices and quantum runtime $T=2\pi N^2$.
    The inset shows the in-constraint probability $\bk{P_\feas}$ over time for the same simulations.
    (\textbf{Middle panels}) Values at the end of simulations with the same random graphs, but different quantum runtimes $T$.
    (\textbf{Right-most panels}) Values at the end of simulations with different qubit numbers (vertices).
    Lines in the right-most panels track average values for each qubit number $N$.
  }
  \label{fig:MIS-summary}
\end{figure*}

We first consider the problem of finding a maximum independent set (MIS).
Given a graph $G=(V,E)$, a set $S \subset V$ is an \emph{independent set} if none of its vertices are neighbors; that is, $\set{v,w}\not\in E$ for all $v,w\in S$.
A \emph{maximum} independent set is an independent set containing the largest possible number of vertices.
Candidate solutions $S\subset V$ to MIS can be identified by bitstrings $x = \set{x_v: v\in V} \in\set{0,1}^{\abs{V}}$ for which $x_v=1$ if and only if $v\in S$.
We refer to identification of subsets as the membership encoding into $\abs{V}$-bit strings.
MIS can thus be defined by the following integer program:
\begin{align}
  \begin{aligned}
    \text{maximize}: ~   & \sum_{v \in V} x_v                           \\
    \text{subject to}: ~ & x_v x_w = 0 ~\text{for all}~\set{v,w} \in E, \\
    & \bm x \in \set{0,1}^{\abs{V}}.
  \end{aligned}
  \label{eq:MIS}
\end{align}
The constraints of MIS are sometimes equivalently written as $x_v + x_w \leq 1$, but the formulation with equality constraints in Eq.~\eqref{eq:MIS} avoids the need for slack variables.
The worst feasible solution to MIS is the empty set, identified by the all-0 bitstring $\bm x = (0, 0, \ldots)$.
Many constrained optimization problems, including as \emph{maximum clique}, \emph{minimum vertex cover}, and \emph{maximum set packing}, are straightforwardly reducible to MIS with no significant overhead \cite{hadfield2019quantum}.

Solving MIS with an adiabatic quantum algorithm was previously considered in the work that inspired Q-CHOP \cite{yu2021quantum}.
Indeed, we show in Appendix \ref{sec:equiv} that solving MIS with Q-CHOP is equivalent to running the algorithm in Ref.~\cite{yu2021quantum} with $\lambda=-4$ and $T=4N^2$.
However, the algorithm in Ref.~\cite{yu2021quantum} provides little insight into the choice $\lambda$, which is identified with $-\d_t\varphi/4 = \lambda^{-1}$ in their work.
As a specific example, Ref.~\cite{yu2021quantum} does not discuss the fact that making $\abs{\d_t\varphi}$ too large ($\abs{\lambda}$ too small) makes the algorithm for MIS fail due to population leakage outside of the feasible subspace.
In contrast, the formalism of Q-CHOP makes clear from Eq.~\eqref{eq:p_feas} that one should choose $\abs{\lambda} \sim N$.
The Q-CHOP formalism additionally clarifies that choosing $\d_t\varphi>0$ ($\lambda<0$) makes these algorithms track the state of \emph{maximal} energy within the feasible subspace, making them more vulnerable to mixing with non-feasible states.
Finally, while Ref.~\cite{yu2021quantum} simulated their algorithm for MIS, they provided no comparison to alternative quantum algorithms.

To evaluate Q-CHOP on MIS and to compare its performance to that of SQAA, we consider random (Erdos-R\'enyi) graphs $G(N,p)$ with $N$ vertices and edge-creation probability $p=0.3$.
Figure \ref{fig:MIS-summary} summarizes simulation results from solving MIS with Q-CHOP and the SQAA on random graphs for different qubit numbers $N$ and quantum runtimes $T$.
The primary takeaway from these results is that Q-CHOP not only outperforms the SQAA in finding high-quality solutions on average (measured by the in-constraint approximation ratio $\bk{r}$), but also finds the optimal solution with significantly higher probability (measured by the projector $P_\opt$).
The probability of finding an optimal solution also grows much faster with respect to the quantum runtime $T$ when solving MIS with Q-CHOP as opposed to the SQAA.
Moreover, for Q-CHOP the probability of finding an optimal solution remains roughly constant with respect to $N$ when $T\sim N^2$, whereas for the SQAA this probability decays with $N$.
For the SQAA, a linear regression test rejects the null hypothesis of constant $\bk{P_\opt}$ with respect to $N$ (at $T=2\pi N^2$) with a $p$-value of $6\times10^{-4}$ (in favor of a negative slope), whereas for Q-CHOP, the null hypothesis is rejected with $p$-value of $0.01$ in favor of a positive slope.
While the utility of a linear regression test is limited for a quantity that is bounded on the interval $[0, 1]$, it illustrates the statistical significance of the trend observed in Figure \ref{fig:MIS-summary}.
Definitively disambiguating this trend from a finite-size effect requires more extensive numerical investigation, and may provide an interesting opportunity for finding asymptotic separations between Q-CHOP and SQAA performance.

\subsection{Directed minimum dominating set}
\label{sec:DMDS}

\begin{figure*}
  \centering
  \includegraphics[scale=0.75]{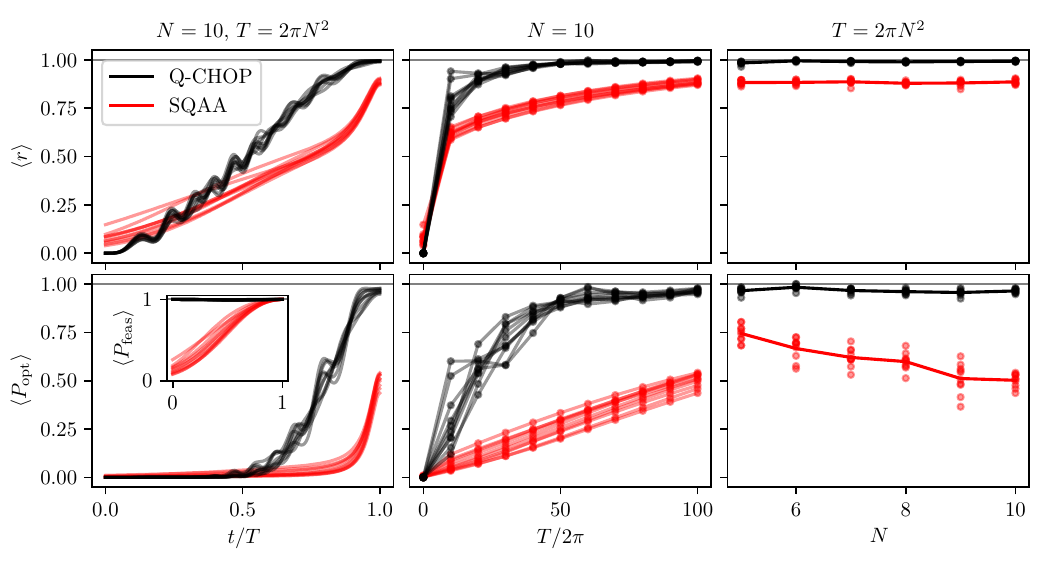}
  \caption{Summary of DMDS simulation results with random directed graphs, presented in a format identical to Figure \ref{fig:MIS-summary}.}
  \label{fig:DMDS-summary}
\end{figure*}

We next consider the problem of finding a directed minimum dominating set (DMDS).
Given a directed graph $G = (V, E)$, a vertex $v\in V$ is \emph{dominated} by a vertex $u\in V$ if the directed edge $(u, v)\in E$.
A \emph{dominating set} of $G$ is then a subset $S\subset V$ for which every vertex of the graph is either contained in $S$, or dominated by a member of $S$.
A \emph{minimum} dominating set of a graph is a dominating set containing the fewest possible vertices.
DMDS can be defined by the following integer program:
\begin{align}
  \begin{aligned}
    \text{minimize}: ~   & \sum_{v \in V} x_v                                                \\
    \text{subject to}: ~ & \bar x_v \prod_{(u,v)\in E} \bar x_u = 0 ~\text{for all}~v \in V, \\
    & \bm x \in \set{0,1}^{\abs{V}},
  \end{aligned}
  \label{eq:DMDS}
\end{align}
where the binary varibale $x_v$ denotes membership of $v$ in $S$, and $\bar x_v = 1 - x_v$ for shorthand.
The constraint for vertex $v$ ensures that either $v\in S$ (in which case $\bar x_v=0$), or $u\in S$ for a dominating neighbor $u:(u,v)\in E$ (in which case $\bar x_u=0$).
The worst feasible solution to DMDS is the all-inclusive set $S=V$, identified by the all-1 bitstring $\bm x=(1, 1, \cdots)$.
The integer program formulation of DMDS in Eq.~\eqref{eq:DMDS} superficially resembles that of MIS in Eq.~\eqref{eq:MIS}, and indeed the problem of finding a minimum dominating set on an \emph{undirected} graph can be reduced to solving MIS on a hypergraph.
The definition of DMDS on a directed graph, however, prevents its straightforward reduction to a variant of MIS.

Figure \ref{fig:DMDS-summary} provides a summary of using Q-CHOP and the SQAA to solve DMDS on directed Erdos-R\'enyi graphs with edge-creation probability $p=0.3$, where existing edges are assigned a random orientation.
Altogether, the results and conclusions from Figure \ref{fig:DMDS-summary} are strikingly similar to those for MIS, likely due to the similarity of the underlying ensemble of graphs that define our MIS and DMDS problem instances.
As with MIS, Q-CHOP yields high-quality solutions to DMDS on average, and has a higher probability of finding an optimal solution.
The optimal-state probability for Q-CHOP again grows faster with quantum runtime $T$, and does not decay with problem size $N$ (Q-CHOP data in the is consistent with the null hypothesis of no decay, while for the SQAA the null hypothesis is rejected with $p$-value $5\times 10^{-4}$), suggesting a potential advantage in scaling that warrants further investigation.

\subsection{Knapsack}
\label{sec:KS}

\begin{figure*}
  \centering
  \includegraphics[scale=0.75]{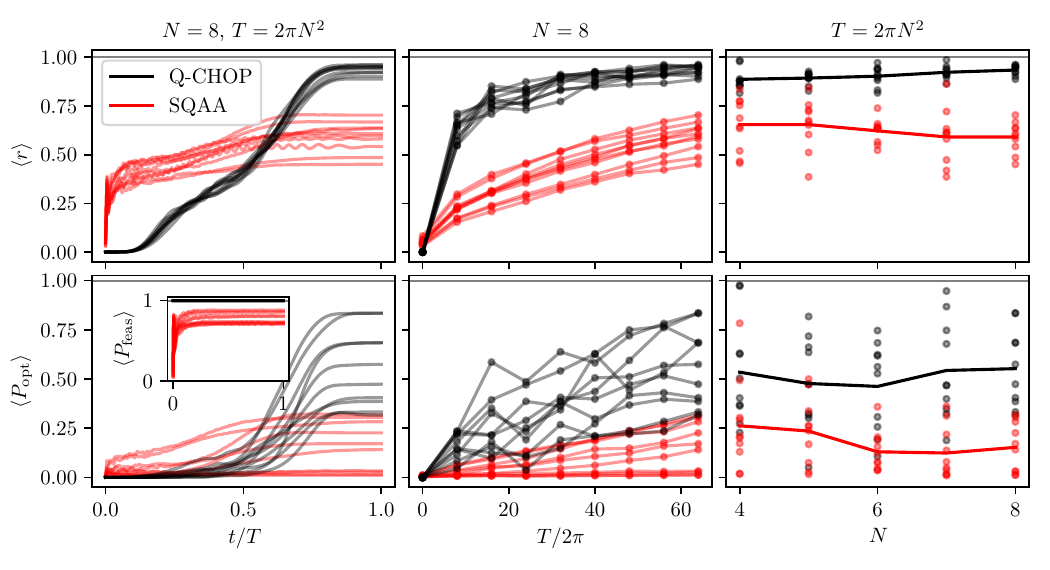}
  \caption{
    Summary of knapsack simulation results, presented in a format identical to Figure \ref{fig:MIS-summary}.
    Here $N$ is the number of items in a knapsack instance.
  }
  \label{fig:KS-summary}
\end{figure*}

\begin{figure*}
  \centering
  \includegraphics[scale=0.75]{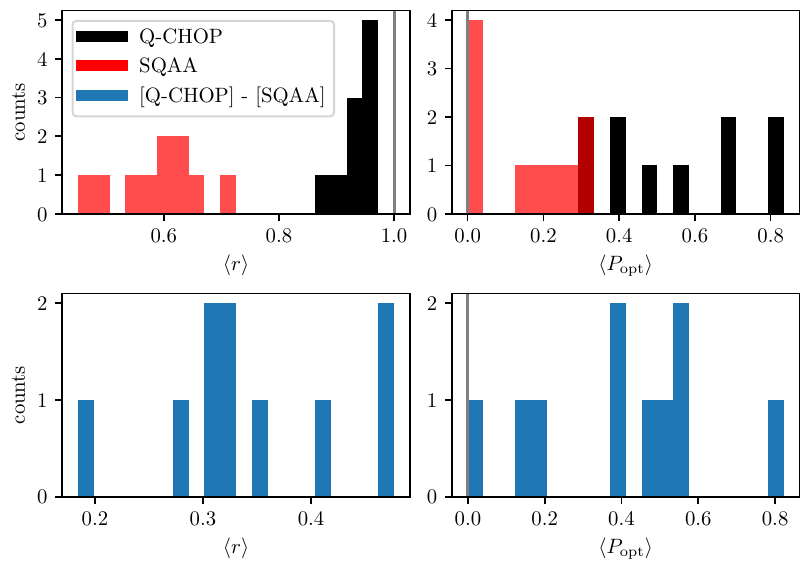}
  \caption{
    \textbf{(Top panels)} Histograms of final in-constraint approximation ratio $\bk{r}$ and optimal-state probability $\bk{P_\opt}$ from knapsack simulations with $N=8$ items and quantum runtime $T=2\pi N^2$.
    The data in the top panels here is a coarse-grained slice of the data at time $t=T$ in the left-most panels of Figure \ref{fig:KS-summary}.
    Darker red bars indicate that Q-CHOP and SQAA values lie within the same histogram bin.
    \textbf{(Bottom panels)} Histograms of the performance differences (as measured by $\bk{r}$ and $\bk{P_\opt}$) between Q-CHOP and the SQAA in the same simulations as the top panels.
    Q-CHOP outperforms the SQAA in every simulated knapsack problem instance with $N=8$ items.
  }
  \label{fig:KS-histograms}
\end{figure*}

We now go beyond graph problems and consider the knapsack problem, a widely studied combinatorial optimization problem \cite{kellerer2004multidimensional, cacchiani2022knapsack} with applications in logistics \cite{kolhe2010planning, perboli2014multihandler, lin2017modeling, verma2020generalized} and finance \cite{vaezi2019portfolio, medvedeva2021randomized}.
A knapsack problem is specified by a collection of items $I$ to be placed in a ``knapsack''.
Each item $i \in I$ has a weight $w_i$ and value $v_i$, and the knapsack has weight capacity $W$.
The goal is to choose a subset of items that maximizes value without exceeding the capacity of the knapsack.
The knapsack problem can be defined by the following integer program:
\begin{align}
  \begin{aligned}
    \text{maximize}: ~ & \sum_{i\in I} v_i x_i          \\
    \text{subject to}: ~     & \sum_{i\in I} w_i x_i \leq W,  \\
    & \bm x \in \set{0,1}^{\abs{I}}.
  \end{aligned}
  \label{eq:KS}
\end{align}
Solving the knapsack problem requires the enforcement of inequality constraints.
Simulating Q-CHOP and the SQAA for a knapsack problem instance therefore requires a total Hilbert space dimension of $2^{\abs{I}} (W+1)$.
The worst feasible solution of a knapsack problem is the all-$0$ state.

The task of generating classically hard knapsack problem instances was previously studied in Ref.~\cite{jooken2022new}, with an implementation of their instance generator provided at Ref.~\cite{jorikjooken2024knapsack}.
In addition to a number of items $N$ and knapsack capacity $W$, an ensemble of random instances generated by the method of Ref.~\cite{jooken2022new} is defined by a number of groups $g$ that determine item weight and value statistics, a fraction of items $f$ that are assigned to the last group, an upper bound $s$ on the value and weight of items in the last group, and a noise parameter $\epsilon$.
To produce random knapsack problem instances with $N$ items, we use the generator in Ref.~\cite{jorikjooken2024knapsack} with $(W, g, f, s, \epsilon) = (2N, \lceil N/2\rceil, 1/N, N, 0.1)$.
We reject knapsack problem instances in which any item has weight greater than $W$.

Figure \ref{fig:KS-summary} summarizes the results of our knapsack problem simulations with Q-CHOP and the SQAA, and additional data about results at $N=8$ and $T=2\pi N^2$ is provided in Figure \ref{fig:KS-histograms}.
The performance of Q-CHOP and the SQAA for the knapsack problem is more variable than that for MIS and DMDS, which can be attributed to a richer ensemble of problem instances.
The SQAA generally fails to produce final states that lie within the feasible-state manifold, and in some cases essentially fails to find an optimal state altogether.
Interestingly, both Q-CHOP and the SQAA exhibit plateau behavior in the growth of $\bk{r}$ and $\bk{P_\opt}$ with increasing progress $t/T$ throughout the solution of individual problem instances.
Even so, Q-CHOP achieves reliably high in-constraint approximation ratios, and finds optimal states with higher probability than the SQAA in every simulated instance (see Figure \ref{fig:KS-histograms}).
Q-CHOP also achieves significantly higher in-constraint approximation ratios and optimal-state probabilities with shorter quantum runtimes $T$ than the SQAA, though there is high variability in the growth of $\bk{P_\opt}$ with $T$.
The performance gap between Q-CHOP and the SQAA persists for different problem sizes $N$.

\subsection{Combinatorial auction}
\label{sec:CA}

\begin{figure*}
  \centering
  \includegraphics[scale=0.75]{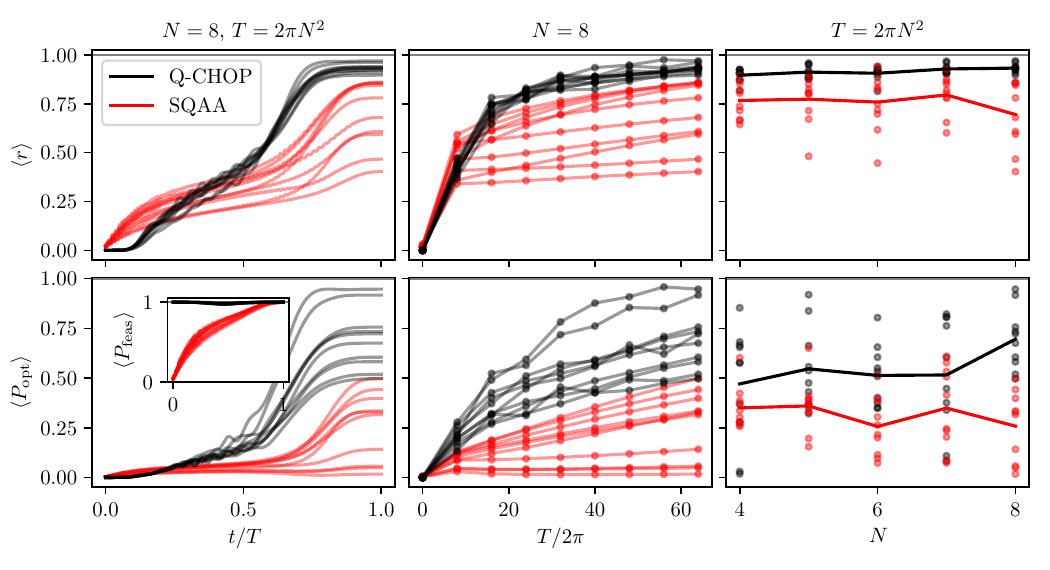}
  \caption{
    Summary of combinatorial auction simulation results, presented in a format identical to Figure \ref{fig:MIS-summary}.
    Here $N$ is the number of bids in the combinatorial auction.
  }
  \label{fig:CA-summary}
\end{figure*}

\begin{figure*}
  \centering
  \includegraphics[scale=0.75]{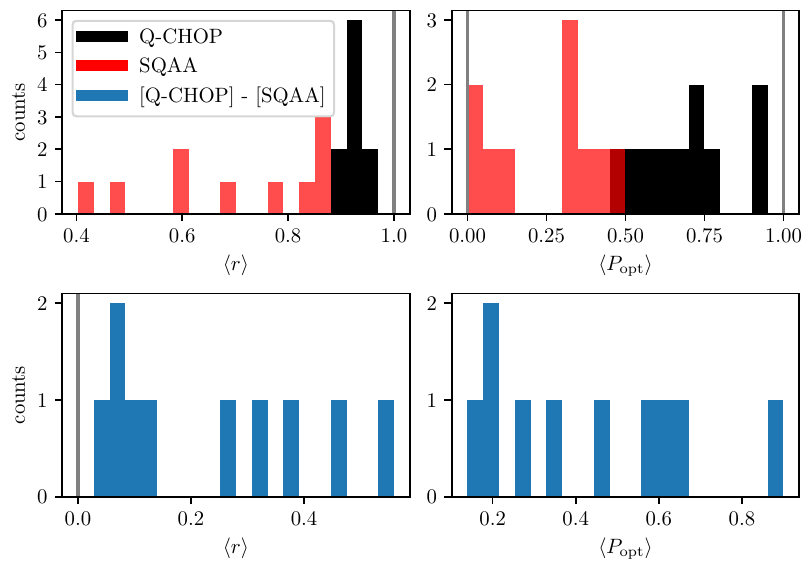}
  \caption{
    Histograms of combinatorial simulation results for $N=8$ bids and quantum runtime $T=2\pi N^2$, presented in a format identical to Figure \ref{fig:KS-histograms}.
    Q-CHOP outperforms the SQAA in every simulated combinatorial auction problem instance with $N=8$ bids.
  }
  \label{fig:CA-histograms}
\end{figure*}

We next consider the problem of optimizing a combinatorial auction (CA), which is a commonly used mechanism for market-based resource allocation and procurement \cite{leyton-brown2000universal, devries2003combinatorial, zhu2016virtualization, ausubel2017practical, prasad2018combinatorial, kayal2019distributed, zhong2020multiresource}.
A combinatorial auction problem instance is specified by a collection $I$ of items to be sold, and a set $B$ of bids that offer to purchase baskets of items.
Each item $i\in I$ has multiplicity $m_i$.
Each bid $b = (p_b, \bm q_b)\in B$ offers to pay $p_b$ for the basket of items specified by the vector $\bm q_b = (q_{bi} :i\in I)$, where $q_{bi}$ is the quantity of item $i$ in the basket requested by bid $b$.
The goal of a combinatorial auction is for the seller to accept a subset of the bids that maximizes payments without exceeding the available inventory of items.
The combinatorial auction problem can be defined by the following integer program:
\begin{align}
  \begin{aligned}
    \text{maximize}: ~   & \sum_{b\in B} p_b x_b                                     \\
    \text{subject to}: ~ & \sum_{b\in B} q_{bi} x_b \leq m_i ~\text{for all}~i\in I, \\
    & \bm x\in\set{0,1}^{\abs{B}}.
  \end{aligned}
  \label{eq:CA}
\end{align}
Due to the inequality constraints and associated slack variables, simulating a combinatorial auction problem instance with Q-CHOP or the SQAA requires a total Hilbert space dimension of $2^{\abs{I}} \prod_{i\in I} (m_i+1)$.
The worst feasible solution of the combinatorial auction is the all-$0$ state.

As can be seen by comparing Eq.~\eqref{eq:KS} to Eq.~\eqref{eq:CA}, the knapsack problem can be interpreted as a combinatorial auction consisting of one item with multiplicity $W$.
A combinatorial auction of unique items (in which all multiplicities $m_i=1$) can be also be reduced to an instance of \emph{weighted} MIS on a graph $(V, E)$ in which nodes are identified with bids, $V = B$, and weighted by the payments $p_b$ in the objective function, as in Eq.~\eqref{eq:CA}.
The edges $E$ are obtained by inserting, for each item $i\in I$, a complete graph on the bids that request item $i$.
In other words, $E = \bigcup_{i\in I} E_i$ with $E_i = \set{\set{a,b}: q_{ai} = q_{bi} = 1}$.

To generate combinatorial auction problem instances, we use the implementation of the Combinatorial Auction Test Suite (CATS) \cite{leyton-brown2000universal} provided by Ref.~\cite{gasse2019exact}.
We thus generate instances of combinatorial auctions with $\abs{I}=3$ items and $N$ bids for varying $N$.
All other parameters defining an ensemble of combinatorial auctions are set to the defaults in Ref.~\cite{gasse2019exact}.
CATS problem instances have all multiplicities $m_i=1$, so our combinatorial auction instances are equivalent to instances of weighted MIS drawn from an appropriately defined ensemble of weighted graphs.
To test Q-CHOP and the SQAA on problems with inequality constraints, however, we will keep the formulation of the combinatorial auction in Eq.~\eqref{eq:CA}, which is simulated with auxiliary slack variables.

Figures \ref{fig:CA-summary} and \ref{fig:CA-histograms} summarize our combinatorial auction simulation results.
Similarly to the results for the knapsack problem, the performance of both Q-CHOP and SQAA is variable across different problem instances.
Despite exhibiting plateau behavior, Q-CHOP still outperforms the SQAA, achieving reliably higher in-constraint approximation ratios, higher optimal-state probabilities, and a faster growth of these performance metrics with quantum runtime $T$.
The performance gap between Q-CHOP and the SQAA persists for different problem sizes $N$.

\subsection{Bond exchange-traded fund basket optimization}
\label{sec:ETF}

\begin{figure*}
  \centering
  \includegraphics[scale=0.75]{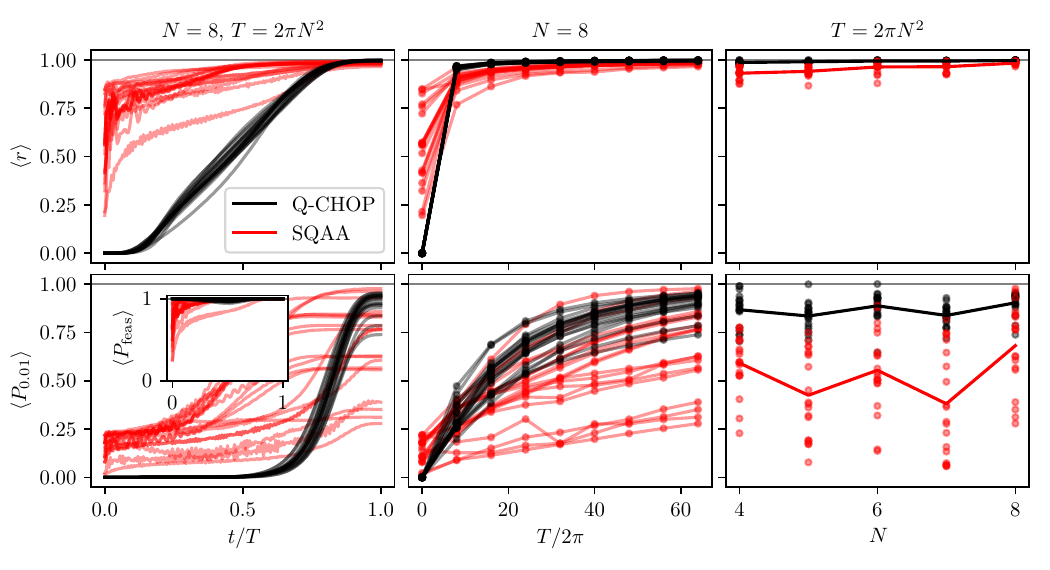}
  \caption{
    Summary of simulation results for 20 bond ETF problem instances, presented in a format identical to Figure \ref{fig:MIS-summary}.
    Here $N$ is the number of assets in an bond ETF problem instance.
  }
  \label{fig:ETF-summary}
\end{figure*}

\begin{figure*}
  \centering
  \includegraphics[scale=0.75]{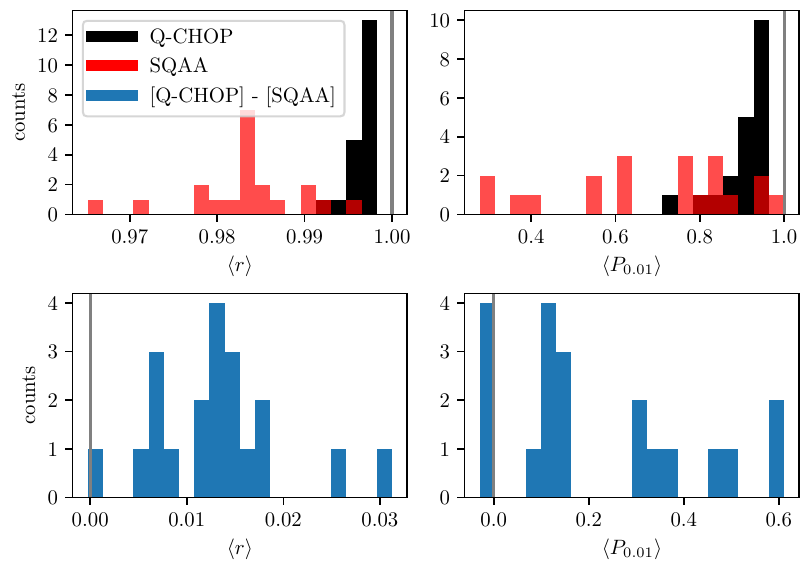}
  \caption{
    Histograms of bond ETF simulation results for $N=8$ assets and quantum runtime $T=2\pi N^2$, presented in a format identical to Figure \ref{fig:KS-histograms}.
    Q-CHOP outperforms the SQAA in all simulated instances as measured by the in-constraint approximation ratio $\bk{r}$, and in 16 of 20 (80\%) of simulated instances as measured by the 0.01-optimal-state probability $\bk{P_{0.01}}$, although the SQAA outperforms Q-CHOP on $\bk{P_{0.01}}$ by at most $0.03$, whereas Q-CHOP outperforms the SQAA by $0.22$ on average across all simulated instances.
  }
  \label{fig:ETF-histograms}
\end{figure*}

Finally, we consider an optimization problem that arises in a financial use case, namely the optimization of bond exchange-traded fund (ETF) basket proposition.
An ETF is an investment vehicle that allows retail and institutional investors to gain exposure to a (typically, large) collection of assets.
To accomplish this, an ETF sponsor issues shares that are tradable on an exchange.
The price of ETF shares is kept in line with the value of the underlying assets through a creation-redemption mechanism.

Denote the net asset value (NAV) of the collection of assets underlying the ETF by
\begin{align}
  \NAV(\ETF) = \sum_a w_a \Price(a),
\end{align}
where $w_a$ is the weight (relative abundance) of asset $a$ in the ETF, with $\sum_a w_a = 1$.
Upon issuing shares of the ETF, the actual price of an ETF share, $\Price(\ETF)$, can fluctuate freely from the net asset value.
To stabilize share prices to the net asset value, authorized participants can create or redeem ETF shares.
If $\Price(\ETF) < \NAV(\ETF)$, authorized participants can redeem ETF shares with the ETF issuer in exchange for a fraction of the ETF assets, and sell these assets for a premium on the secondary market.
Similarly, if $\Price(\ETF) > \NAV(\ETF)$, authorized participants can buy a basket of assets on the secondary market, exchange this basket with the ETF issuer for newly issued ETF shares, and sell these shares for a premium on the secondary market.
In both cases, the authorized participant profits from a difference between the net asset value and the market price of an ETF share, provided that the authorized participant can identify a suitable basket of assets to exchange with the ETF issuer.

Unlike stocks, bonds typically trade in discrete increments.
The discrete nature of bond trading makes the problem of bond ETF basket proposition naturally discrete and thereby more challenging to optimize.

Consider the task of designing a basket of assets that can be exchanged for bond ETF shares.
The market price of a basket $B_{\bm x}$ represented by the asset vector $\bm x$ is
\begin{align}
  \Price(B_{\bm x}) = \sum_a \Price(a) x_a,
\end{align}
where $x_a$ indicates the quantity of bond $a$ that is included in the basket, and for simplicity we restrict all $x_a\in\set{0, 1}$, such that altogether $\bm x\in\set{0, 1}^N$.
To preserve the characteristics of the bond ETF portfolio (for example, to track a specified index like the S\&P 500 Bond Index), the bond ETF issuer defines a set of constraints on baskets that may be exchanged for bond ETF shares.
These constraints often take the form
\begin{align}
  \ell \sum_a x_a \le \sum_{a\in F} x_a \le u \sum_a x_a,
  \label{eq:ETF_cons}
\end{align}
where $F$ can represent (for example) a market sector (such as financial, consumer electronics, etc.), and $\ell,u$ are real numbers that bound the bond ETF exposure to $F$.
In practice, strict constraints can make it too difficult to design constraint-satisfying baskets, so the bond ETF issuer may relax the hard requirements for a basket corresponding to an integer number of bond ETF shares (for example, by decreasing $\ell$ or increasing $u$) in favor of a ``soft'' penalty that discourages an exchange-value mismatch
\begin{align}
  \Delta(\bm x) = m \NAV(\ETF) - \Price(B_{\bm x})
\end{align}
that needs to be resolved in cash.
Here $m$ is the number of bond ETF shares exchanged for the basket.
Specifically, the penalty resolving a mismatch in cash takes the form of a fee,
\begin{align}
  f(\bm x) = g \Delta(\bm x)^2,
\end{align}
that the exchanging party wishes to minimize.
Here $g$ is a penalty factor chosen by the bond ETF issuer.
Altogether, the optimization problem is to identify an asset vector $\bm x$ that minimizes the fee $f(\bm x)$ while satisfying constraints of the form in Eq.~\eqref{eq:ETF_cons}.

To generate bond ETF problem instances with $N$ assets, we set the number of bond ETF shares to $m = \lceil N/2+1 \rceil$, which is sufficiently high to ensure that the all-0 vector corresponding to an empty basket is the worst feasible solution.
We then sample bond ETF asset weights $w_a$ uniformly on the interval $(0, 1)$ before normalizing to ensure that $\sum_a w_a = 1$, and sample the asset prices $\Price(a)$ from a normal distribution with mean $1$ and standard deviation $0.1$.
These quantities determine the objective function $f(\bm x)\propto\Delta(\bm x)^2$ up to an arbitrary coefficient.
To determine the constraints, we assign each asset one of three market sectors $F_1,F_2,F_3$ uniformly at random.
The induced constraints of the form in Eq.~\eqref{eq:ETF_cons} are then determined by
\begin{align}
  \ell_j = (1 - \epsilon) \sum_{a\in F_j} w_a,
  &&
  u_j = (1 + \epsilon) \sum_{a\in F_j} w_a,
\end{align}
where we set $\epsilon = 0.1$.

To ensure that inequality constraints are integer-valued (which is necessary for their simulation via the slack variable method in Section \ref{sec:inequality}), we multiply the inequalities in Eq.~\eqref{eq:ETF_cons} by a factor of 10 and round coefficients to the nearest integer.
In practice, the high dimensionality of the slack variables necessary to enforce all inequality constraints makes their classical simulation computationally intractable, so we enforce only one upper bound constraint for each simulated problem instance.
We also reject instances in which all states are feasible, or in which all feasible states have the same objective energy (leading to an undefined approximation ratio).

Figures \ref{fig:ETF-summary} and \ref{fig:ETF-histograms} summarize our bond ETF simulation results.
Both Q-CHOP and the SQAA perform well as measured by the in-constraint approximation ratio.
However, Q-CHOP is more likely to produce a solution with very high in-constraint approximation ratio, and on average there is a significant performance gap between Q-CHOP and the SQAA in the probability of finding a near-optimal solution (now relaxed to being 0.01-optimal because objective function values are not integer-valued, and may therefore be arbitrarily close together).
There is no longer a clean separation between Q-CHOP and the SQAA in how their near-optimal-state probability grows with the quantum runtime $T$, but typical growth for Q-CHOP is more favorable than that for the SQAA.

\section{Discussion and future directions}
\label{sec:discussion}

Q-CHOP is a new adiabatic quantum algorithm for solving constrained optimization problems.
Whereas the standard approach to incorporating constraints is to first modify the objective to penalize constraint violation, and then proceed with an ordinary adiabatic algorithm for unconstrained optimization, Q-CHOP enforces constraints throughout evolution, and uses the objective function to construct an adiabatic path that converts the worst feasible state into the best feasible state.
Q-CHOP thereby assigns constraint and objective function qualitatively distinct roles, which futher motivates a natural choice for the constraint-enforcing ``penalty factor'' $\lambda$, although additional hyperparameter optimization (or a more refined theory) would yield even better choices of $\lambda$.
Q-CHOP consistently outperforms the standard, penalty-based adiabatic algorithm in numerical experiments across a diverse set of problems, suggesting that Q-CHOP is a promising general-purpose method for constrained combinatorial optimization.

It is worth mentionining that our conclusions about how Q-CHOP compares to the SQAA is contingent on the choice of penalty factor $\lambda$.
While we chose to keep $\lambda$ constant for both methods, it is possible that optimizing over $\lambda$ may diminish the performance difference between Q-CHOP and the SQAA.
In a similar vein, there are additional optimization parameters in the SQAA, namely the choice of driver Hamiltonian and its scalar prefactor, which also affect its performance.
Both methods also have freedom in their choice of adiabatic schedule.
Future work could therefore examine how Q-CHOP and the SQAA compare when optimizing over these additional parameters.
We note, however, that the lack of a requirement to perform additional hyperparameter optimization is a notable advantage of Q-CHOP.

There remain many important open questions about Q-CHOP that could inspire exciting future research directions.
First, the design of Q-CHOP involved several choices that may be worth examining more closely.
For example, while the strategy to ``rotate'' the objective Hamiltonian was naturally obtained by generalizing the physically-inspired specialized algorithm finding independent sets in Refs.~\cite{wu2020quantum, yu2021quantum}, it would be interesting to consider different strategies to ``invert'' objective Hamiltonians, which amounts to finding different adiabatic paths.
Second, there is also a clear need to design better strategies to handle inequality constraints, since the strategy in Section \ref{sec:inequality} does not scale to large numbers of inequality constraints.
Third, while having an easy-to-prepare worst feasible state is a common feature of constrained optimization problems, it can also be the case that the worst feasible state is just as difficult to identify as the best feasible state.
It is therefore important to test, validate, and refine our proposed relaxation of the Q-CHOP (in Section \ref{sec:relaxation}) from requiring an easy-to-prepare \emph{worst} feasible state to requiring an \emph{arbitrary} easy-to-prepare feasible state.

More broadly, it is desirable to build a better theoretical understanding of Q-CHOP, and ideally obtain quantitative guarantees or bounds on performance.
As the performance of any adiabatic algorithm is ultimately tied to the minimal spectral gap of its Hamiltonian, bounds and guarantees will depend on the properties of the constraint and objective Hamiltonians.
As an example, Ref.~\cite{yu2021quantum} studied the phenomenon of feasible-subspace ``shattering'' for an MIS problem on an adversarially chosen, loosely connected graph.
This analysis suggests a connection between the performance of Q-CHOP and the connectivity of its feasible-state space when interpreting the rotated objective Hamiltonian as an adjacency matrix.
In the case of a disconnected feasible-state subspace, Q-CHOP may be unable to produce optimal---or even good---solutions entirely, if these solutions lie in different connected components than that of the worst feasible state.
This may occur, for example, if symmetries of the objective Hamiltonian fragment Hilbert space into uncoupled sectors, or if the structure of an optimization problem requires decision variables to change in a correlated manner for one feasible state to change into another
\footnote{
  As a concrete example, the \emph{quadratic assignment problem} optimizes over the space of permutation matrices.
  If these matrices are represented as $X=\sum_{i,j} x_{ij} \op{i}{j}$ with binary variables $x_{ij}\in\set{0,1}$, then the induced constraints $\sum_i x_{ij} = 1$ and $\sum_j x_{ij} = 1$ require at least \emph{four} decision variables to change concurrently to get from one feasible state to another.
}.
It is therefore important to develop a better understanding of when and why loosely connected or disconnected subspaces can occur, both to understand the limitations of Q-CHOP, and to design strategies to mitigate these performance obstacles.

In addition to subspace shattering, other leads for developing a better theoretical understanding Q-CHOP include the enticing observation that Q-CHOP exhibits ``universal'' behavior for the dynamics of the in-constraint approximation ratio $\bk{r}$ with respect to progress $t/T$ within each problem class that was considered in Section \ref{sec:numerics}.
While the SQAA exhibits similar universal behavior for MIS and DMDS (Figures \ref{fig:MIS-summary} and \ref{fig:DMDS-summary}), which can likely be understood by studying the statistics of Erdos-R\'enyi graphs, Q-CHOP continues to exhibit this behavior for problems with considerably more structure (Figures \ref{fig:KS-summary}, \ref{fig:CA-summary}, and \ref{fig:ETF-summary}).

Given the sparse literature on constrained optimization with quantum adiabatic algorithms, it would also be interesting to borrow ideas from Q-CHOP to modify algorithms other than the SQAA.
Reverse annealing algorithms \cite{perdomo-ortiz2011study, callison2022hybrid}, for example, may similarly be augmented to enforce constraints at all times with a natural choice of penalty factor.

Finally, it would be interesting to study a discretized, variational variant of Q-CHOP, in direct analogy to the quantum alternating operator ansatz (QAOA) \cite{farhi2014quantum, hadfield2019quantum} that is obtained by discretizing the SQAA.
Recent work \cite{shaydulin2024evidence} has shown the first evidence of an asymptotic scaling advantage for QAOA on a classically hard problems, and a recursive variant known as RQAOA can be shown to further improve QAOA performance \cite{bravyi2020obstacles, bravyi2022hybrid, bae2024recursive}.
The performance gap observed between Q-CHOP and the SQAA in our work suggests that a discretized variant of Q-CHOP---and derivatives thereof---have the potential to further improve on the performance of QAOA and RQAOA.


\bibliography{main.bib}

\section*{Disclaimer}
This paper was prepared for informational purposes with contributions from the Global Technology Applied Research center of JPMorgan Chase \& Co. This paper is not a product of the Research Department of JPMorgan Chase \& Co.~or its affiliates. Neither JPMorgan Chase \& Co.~nor any of its affiliates makes any explicit or implied representation or warranty and none of them accept any liability in connection with this position paper, including, without limitation, with respect to the completeness, accuracy, or reliability of the information contained herein and the potential legal, compliance, tax, or accounting effects thereof. This document is not intended as investment research or investment advice, or as a recommendation, offer, or solicitation for the purchase or sale of any security, financial instrument, financial product or service, or to be used in any way for evaluating the merits of participating in any transaction.

\appendix

\section{Equivalence of Q-CHOP to the method of Yu, Wilczek, and Wu}
\label{sec:equiv}

Here we demonstrate the equivalence of Q-CHOP to the specialized method for finding independent sets in Refs.~\cite{wu2020quantum, yu2021quantum}.
The objective and constraint Hamiltonians for MIS defined on a graph $G = (V, E)$ are
\begin{align}
  H_\obj^\MIS &= \sum_{v\in V} \op{1}_v \simeq S_z, \label{eq:H_obj_MIS} \\
  H_\con^\MIS &= \sum_{\set{v,w}\in E} \op{11}_{v,w}.
  \label{eq:H_con_MIS}
\end{align}
The algorithm in Ref.~\cite{yu2021quantum} was essentially to evolve an initially all-zero state $\ket{00\cdots0}$ under the rotating constraint Hamiltonian
\footnote{
  For reference, note that there is a sign difference between our definition of the Pauli-$Z$ operator and that in Ref \cite{yu2021quantum}, and that our definition of $V$ in Eq.~\eqref{eq:H_con_rot} differs from that in Ref.~\cite{yu2021quantum} by an (inconsequential) gauge transformation.
}
\begin{align}
  H_\con^\rot = V H_\con^\MIS V^\dag,
  &&
  V = R_\z(\phi)^\dag R_\y(\theta)^\dag,
  \label{eq:H_con_rot}
\end{align}
where $R_\alpha(\eta) = e^{-\ii\eta S_\alpha}$ rotates all qubits about axis $\alpha$ by $\eta$, and the angles $(\theta, \phi)$ are functions of time, with $\theta(0) = \phi(0) = 0$.
The unitary generated by evolving under $H_\con^\rot$ up to time $T$ at which $\theta=\pi$ is
\begin{align}
  U_\feas^\rot(T) = \prod_{t=0}^T e^{-\ii \dd t\, H_\con^\rot(t)}
  = \prod_{t=0}^T V(t) e^{-\ii \dd t\, H_\con^\MIS} V(t)^\dag.
\end{align}
We can prepend (on the right) an initial rotation of $V(-\dd t) = 1 + O(\dd t)$ and regroup factors to write
\begin{align}
  U_\feas^\rot(T) = V(T) \prod_{t=0}^T e^{-\ii \dd t\, H_\con^\MIS}
  V(t)^\dag V(t-\dd t),
\end{align}
where
\begin{align}
  V(t)^\dag V(t-\dd t)
  &= V(t)^\dag \sp{V(t) - \dd t \, \d_t V(t)} \\
  &= 1 - \dd t \, V(t)^\dag \d_t V(t) \\
  &= e^{-\ii\dd t G_V(t)},
\end{align}
with
\begin{align}
  G_V = -\ii V^\dag \d_t V
  = \d_t\phi\, R_\y(\theta) S_\z R_\y(\theta)^\dag + \d_t\theta\, S_\y.
\end{align}
We thus find that
\begin{align}
  U_\feas^\rot(T) = V(T) \prod_{t=0}^T e^{-\ii \dd t\, H_\eff(t)},
  \label{eq:U_yu}
\end{align}
where the effective Hamiltonian is
\begin{align}
  H_\eff
  &= H_\con^\MIS + G_V \\
  &=  H_\con^\MIS + \d_t\phi\, R_\y(\theta) S_\z R_\y(\theta)^\dag + \d_t\theta\, S_\y.
  \label{eq:H_eff_yu}
\end{align}
In the limit $\d_t\theta\ll\d_t\phi$, which was enforced in Ref.~\cite{yu2021quantum}, this Hamiltonian becomes
\begin{align}
  H_\eff \approx H_\con^\MIS + \d_t\phi\, R_\y(\theta) S_\z R_\y(\theta)^\dag,
\end{align}
where $S_\z$ is essentially the objective Hamiltonian of MIS, up to an irrelevant scalar term.
The method of Ref.~\cite{yu2021quantum} thus essentially reduces to Q-CHOP, with some minor differences:
\begin{itemize}[leftmargin=*]
  \item The constraint Hamiltonian in Eq.~(2) of Ref.~\cite{yu2021quantum} is greater than that in Eq.~\eqref{eq:H_con_MIS} by a factor of 4 (after setting their energy scale $\Delta=1$)
    \footnote{
      There are also some apparent sign differences between $H_\con^\MIS$ as defined in Ref.~\cite{yu2021quantum} and Eq.~\eqref{eq:H_con_MIS}.
      This is due to a sign difference in our definitions of the Pauli-$Z$ operator.
    }.
    Increasing $H_\con\to 4 H_\con$ is equivalent to taking $(T,\lambda)\to(4T,4\lambda)$ in Q-CHOP.
  \item The ramp in $H_\eff$ above is $\d_t\phi\, R_\y(\theta) H_\obj^\MIS R_\y(\theta)^\dag$, which is equivalent to that in Q-CHOP with $\lambda^{-1}=-\d_t\phi/4$ (after accounting for the factor of 4 mentioned above).
    Assuming $\d_t\phi>0$, the sign difference in prefactors amounts to Ref.~\cite{yu2021quantum} adiabatically tracking the \emph{maximum} energy state within $\H_\feas$, rather than the ground state
    \footnote{
      An implicit choice of $\lambda<0$ is manifest in Eq.~(8) of Ref.~\cite{yu2021quantum}, which time-evolves an initial state under the ``secondary Hamiltonian'' $A$, but with the wrong sign for ordinary time evolution: $\prod_t e^{\ii\dd t A}$ rather than $\prod_t e^{-\ii\dd t A}$.
    }.
    Ref.~\cite{yu2021quantum} set $\d_t\phi=1$ when choosing parameters for simulation, which amounts to choosing $\lambda=-4$.
  \item After time-evolving under the effective Hamiltonian $H_\eff$, the unitary $U_\feas^\rot(T)$ in Eq.~\eqref{eq:U_yu} ends with an application of $V(T) = R_\z(\phi(T))^\dag R_\y(\pi)^\dag$.
    The rotation $R_\z(\phi(T))^\dag$ is diagonal in the computational basis, and therefore has no effect on measurement outcomes.
    The rotation $R_\y(\pi)^\dag$, meanwhile, is accounted for by the fact that Ref.~\cite{yu2021quantum} flips all qubits as an additional step after time evolution, thereby canceling out the effect of $R_\y(\pi)^\dag$.
\end{itemize}
Altogether, the method of Ref.~\cite{yu2021quantum} is thus equivalent to Q-CHOP, with $\lambda=-4$ and $T=4N^2$, where $N$ is the number of vertices in an MIS instance.
Moreover, as we show in Appendix \ref{sec:counterdiabatic}, the additional $\d_t\theta\, S_\y$ term in Eq.~\eqref{eq:H_eff_yu} can be identified with approximate counterdiabatic driving.

\section{Counterdiabatic driving for single-body objectives}
\label{sec:counterdiabatic}

Here we show how Q-CHOP may be sped up for problems with single-body objective Hamiltonians of the form
\begin{align}
  H_\obj = \frac12 \sum_j c_j Z_j,
\end{align}
where $\set{c_j}$ are arbitrary real coefficients.
For convenience and clarity, we define $H_{\obj,\z} = H_\obj$, as well as rotated versions $H_{\obj,\x}$ and $H_{\obj,\y}$ in which the Pauli-$Z$ operators in $H_{\obj,\z}$ are respectively replaced by Pauli-$X$ and Pauli-$Y$ operators.
We will also denote the restriction of Hamiltonians to the feasible subspace $\H_\feas$ by a superscript, such that $H_{\obj,\alpha}^\feas = P_\feas H_{\obj,\alpha} P_\feas$, where $P_\feas$ is a projector onto $\H_\feas$.

Q-CHOP relies on the adiabatic theorem to track the ground state of a Hamiltonian $H(\theta)$ by slowly varying a parameter $\theta$.
Evolving too quickly introduces errors due non-adiabatic effects.
These effects can be mitigated by the introduction of a \emph{counterdiabatic drive} \cite{sels2017minimizing}, in which the Hamiltonian is replaced according to
\begin{align}
  H \to H_\CD = H + \d_t\theta\, G,
\end{align}
where the \emph{gauge field} $G$ can be found by minimizing the Hibert-Schmidt norm of the operator
\begin{align}
  \O(G,H) = \d_\theta H + \ii \sp{G,H}.
\end{align}
Here $\sp{G,H} = GH - HG$, and the (squared) Hilbert-Schmidt norm of $\O$ is $\norm{\O}_{\mathrm{HS}}^2 = \tr(\O^\dag\O)$.

When time evolution is sufficiently slow to avoid exciting states outside the feasible subspace, Q-CHOP effectively evolves from time $t=0$ to $t=T$ under the Hamiltonian
\begin{align}
  H_\tot^\feas(\theta) = -\lambda^{-1} H_\obj^\feas(\theta),
\end{align}
with
\begin{align}
  H_\obj^\feas(\theta) = P_\feas R_\y(\theta)
  H_\obj R_\y(\theta)^\dag P_\feas,
\end{align}
where the angle $\theta:0\to\pi$ as $t:0\to T$.
As $H_\tot^\feas(\theta)$ is the Hamiltonian that we want to augment with a counterdiabatic drive, we want to find a gauge field $G$ that minimizes the norm of $\O(G,H_\tot^\feas)$, which is equivalent to minimizing the norm of $\O(G,H_\obj^\feas)$.
To this end, we expand
\begin{align}
  H_\obj^\feas(\theta) = \cos(\theta) H_{\obj,\z}^\feas + \sin(\theta) H_{\obj,\x}^\feas
\end{align}
and in turn
\begin{align}
  \d_\theta H_\obj^\feas(\theta)
  = -\sin(\theta) H_{\obj,\z}^\feas + \cos(\theta) H_{\obj,\x}^\feas.
  \label{eq:d_theta_H_obj_feas}
\end{align}
It is unclear how to minimize the norm of $\O(G,H_\obj^\feas)$ exactly, so we resort to minimizing it variationally with respect to the ansatz
\begin{align}
  G_g = g_\x S_\x^\feas + g_\y S_\y^\feas + g_\z S_\z^\feas,
\end{align}
where
\begin{align}
  S_\alpha^\feas = P_\feas S_\alpha P_\feas,
\end{align}
for which
\begin{multline}
  \O(G_g, H_\obj^\feas)
  = \d_\theta H_\obj^\feas(\theta)
  + \ii \cos(\theta) \sp{G_g, H_{\obj,\z}^\feas} \\
  + \ii \sin(\theta) \sp{G_g, H_{\obj,\x}^\feas}.
\end{multline}
Nonzero values of $g_\x$ and $g_\z$ only contribute to imaginary components to $\O(G_g,H_\obj^\feas)$, and conversely all imaginary components of $\O(G_g,H_\obj^\feas)$ involve factors of $g_\x$ or $g_\z$.
To minimize the Hilbert-Schmidt norm of $\O(G_g,H_\obj^\feas)$, we therefore set $g_\x=g_\z=0$ and define $G_\y = g_\y S_\y^\feas$, leaving us with
\begin{multline}
  \O(G_\y, H_\obj^\feas)
  = \d_\theta H_\obj^\feas(\theta)
  + \ii g_\y \cos(\theta) \sp{S_\y^\feas, H_{\obj,\z}^\feas} \\
  + \ii g_\y \sin(\theta) \sp{S_\y^\feas, H_{\obj,\x}^\feas}.
  \label{eq:O_G_y}
\end{multline}
The projector $P_\feas$ commutes with $H_{\obj,\z}$ (both are diagonal in the computational basis), which allows us to simplify
\begin{align}
  \sp{S_\y^\feas, H_{\obj,\z}^\feas}
  = P_\feas \sp{S_\y, H_{\obj,\z}} P_\feas
  = \ii H_{\obj,\x}^\feas.
\end{align}
Further substituting $\d_\theta H_\obj^\feas(\theta)$ from Eq.~\eqref{eq:d_theta_H_obj_feas} into Eq.~\eqref{eq:O_G_y}, we thus find that
\begin{multline}
  \O(G_\y, H_\obj^\feas)
  = \cos(\theta) (1-g_\y) H_{\obj,\x}^\feas \\
  - \sin(\theta)
  \p{H_{\obj,\z}^\feas - \ii g_\y \sp{S_\y^\feas, H_{\obj,\x}^\feas}}.
\end{multline}
The $\sin(\theta)$ term vanishes as $\theta\to0$ or $\pi$, at which point $\O(g_\y S_\y^\feas, H_\obj^\feas)$ is minimized by $g_\y=1$.
Near $\theta=\pi/2$, the $\sin(\theta)$ term dominates, but here we can make no definite statements about the norm of $\O(G_\y, H_\obj^\feas)$ without knowledge of the feasible subspace.
Nonetheless, given that $\ii \sp{S_\y, H_{\obj,\x}} = H_{\obj,\z}$, we generally expect $\ii \sp{S_\y^\feas, H_{\obj,\x}^\feas}$ to be ``$H_{\obj,\z}^\feas$-like'', so we still expect a minimum near $g_\y\sim1$.
We conclude that the gauge field $G = S_\y$ is a reasonable (albeit sub-optimal) choice to mitigate non-adiabatic effects.
In the case of linear objectives, the total Hamiltonian for Q-CHOP with approximate counterdiabatic driving thus becomes
\begin{align}
  H_\tot \to H_{\tot,\CD}
  = H_\con - \lambda^{-1} H_\obj(\theta) + \d_t\theta\, S_\y.
\end{align}
The small factor of $\d_t\theta\sim1/T$ prevents $\d_t\theta\,S_\y$ from generating a population transfer to non-feasible states, so $\d_t\theta\,S_\y \approx \d_t\theta\,S_\y^\feas$.

\end{document}